\documentclass[journal=tosc]{iacrtrans}

\usepackage{amsmath}

\usepackage{amssymb,amsfonts}
\usepackage{algorithmic}
\usepackage{graphicx}
\usepackage{textcomp}
\usepackage{standalone}
\usepackage{xcolor}
\def\BibTeX{{\rm B\kern-.05em{\sc i\kern-.025em b}\kern-.08emT\kern-.1667em\lower.7ex\hbox{E}\kern-.125emX}}
\usepackage[utf8]{inputenc}
\usepackage{booktabs} 
\usepackage{graphicx}
\usepackage{subfigure}
\usepackage{subcaption}
\usepackage{units}
\usepackage{blindtext}
\usepackage{multirow}
\usepackage{xcolor,xspace}
\usepackage{amsmath}
\usepackage{paralist}
\usepackage{mdframed}
\usepackage{pict2e}
\usepackage{listings}
\usepackage{color}
\usepackage{threeparttable}
\usepackage{array}
\usepackage{booktabs}
\usepackage{pifont}
\usepackage{url}
\usepackage{comment}
\usepackage{amsmath}
\usepackage{lipsum}
\usepackage{arydshln}
\usepackage{flushend}
\usepackage{latexsym}
\usepackage{enumitem}
\usepackage{fancyhdr}
\usepackage[english]{babel}
\usepackage{tabularx}
\input{glyphtounicode}

\usepackage[ruled]{algorithm2e}
\usepackage{algorithmic}

\usepackage{url}
\usepackage[
    n,
    advantage,
    operators,
    sets,
    adversary,
    landau,
    probability,
    notions,
    logic,
    ff,
    mm,
    primitives,
    events,
    complexity,
    asymptotics,
    keys]{cryptocode}

\usepackage{tikz, pgfplots, pgfplotstable}                    
\usetikzlibrary{arrows, patterns, automata, positioning}
\usepgfplotslibrary{statistics}

\newcommand{\ignore}[1]{}

\newcommand{\app}{{\ensuremath{\sf{App}}}\xspace}
\newcommand{\prv}{{\ensuremath{\sf{\mathcal Prv}}}\xspace}
\newcommand{\vrf}{{\ensuremath{\sf{\mathcal Vrf}}}\xspace}

\newcommand{\chal}{{\textit{Chal}}\xspace}
\newcommand{\RA}{{\textit{RA}}\xspace}

\newcommand{\CFA}{{\textit{CFA}}\xspace}

\newcommand{\acron}{\textit{{RESPEC-CFA}}\xspace}

\newcommand{\attkey}{\ensuremath{\mathcal K}\xspace}

\renewcommand\adv{\ensuremath{\sf{\mathcal Adv}}\xspace}

\newcommand{\cfattest}{\ensuremath{\mathtt{MAC}}\xspace}

\newcommand{\vrfy}{\ensuremath{\mathtt{Verify}}\xspace}

\newcommand{\cflog}{\ensuremath{CF_{Log}}\xspace}
\newcommand{\prefix}{\textit{prefix\ensuremath{_{act}}}\xspace}
\newcommand{\len}{$\mathit{prefix_{len}}$\xspace}

\newcounter{protocol}

\newlist{myenumerate}{enumerate}{1}
\setlist[myenumerate]{
  label=\arabic*),
  align=left,
  leftmargin=*,
  nosep,
}

\newlist{myitemize}{itemize}{1}
\setlist[myitemize]{
    label=\textbullet,
    align=left,
    leftmargin=*,
    nosep,
}

\newcommand*\circled[1]{\tikz[baseline=(char.base)]{
            \node[shape=circle,fill,inner sep=1pt] (char) {\textcolor{white}{#1}};}}

\author{Liam Tyler\inst{1} \and Adam Caulfield\inst{2} \and Ivan De Oliveira Nunes\inst{3}}
\institute{
  University of Zurich, Zurich, Switzerland, \email{ltyler@ifi.uzh.ch}
  \and
  University of Waterloo, Waterloo, Canada, \email{acaulfield@uwaterloo.ca}
  \and  University of Zurich, Zurich, Switzerland, \email{ivan.deoliveiranunes@uzh.ch}
}

\title[Efficient CFA by Speculating on Control Flow Path Representations]{Efficient Control Flow Attestation by Speculating on Control Flow Path Representations}

\begin{document}

\maketitle


\begin{abstract}
Microcontroller Units (MCUs) are ubiquitous and perform safety-critical sensing and actuation within larger cyber-physical systems. Yet, despite their essential role, MCUs have significant resource constraints and often lack the (more robust) architectural security features of general-purpose computers. This leaves them vulnerable to runtime attacks that can remotely modify their code or violate their execution integrity.

To secure MCUs in an affordable fashion, prior work has proposed low-cost methods for remote verification of an MCU's software state. Among them, Remote Attestation (\RA) is a challenge-response protocol wherein a remote Verifier (\vrf) issues a cryptographic challenge and requests a timely response from a potentially compromised Prover MCU (\prv). A root of trust within \prv is responsible for producing evidence of \prv's software state by computing an authenticated integrity check (e.g., a MAC or signature) over the current snapshot of \prv's program memory and the received challenge. By examining the produced response message, \vrf can determine if \prv's code has been illegally modified (e.g., via code injection attacks or by reprogramming the MCU through local/physical interfaces).

Control Flow Attestation (\CFA) schemes augment \RA to also produce an authenticated log of the control flow path taken during the execution of an attested software operation. This allows \vrf to inspect the control flow path to detect and gain visibility of the behavior of control flow attacks, in addition to illegal code modifications. However, an important bottleneck in \CFA is the storage and transmission of control flow logs. To address this, \CFA optimizations have been proposed with state-of-the-art methods focusing on application-specific optimizations that speculate on likely control flow sub-paths. The key idea is to replace likely paths with reserved symbols of reduced size, thereby reducing the overall size of control flow logs without loss of information. 

Despite this progress, we argue that prior approaches overlook the data representation of control flow paths in their speculation strategy. Based on this observation, we propose \acron, an architectural extension for \CFA allowing control flow path speculation based on (1) the locality of control flow paths and (2) their Huffman encoding. \acron alone reduces control flow log sizes by up to 90.1\%. We also strive to design \acron such that it can compose synergistically with state-of-the-art methods. As a result, when combined with prior methods, \acron achieves reductions of up to 99.7\% in log sizes (without loss of information), significantly outperforming previous approaches and advancing practical \CFA.

\end{abstract}

\section{Introduction}
Modern cyber-physical systems depend on Microcontroller Units (MCUs) for sensing and actuation. However, given their low cost and low energy requirements, MCUs often lack security features comparable to general-purpose computers. For example, they typically lack Memory Management Units (MMUs), inter-process isolation, or strong privilege level separation (see Section~\ref{sec:scope} for more details on MCU architectures). Yet, MCUs often perform system-critical tasks as a part of larger systems in which they are embedded, making them attractive targets of attacks~\cite{kayan2022cybersecurity}. Therefore, reliable methods to assess the integrity of remote MCUs are crucial.

Among cost-effective methods for remote integrity verification, Remote Attestation (\RA)~\cite{vrased, Sancus17, sake} is a two-party protocol that allows a Verifier (\vrf) to remotely measure the software state of a remote Prover MCU (\prv). In \RA, \vrf requests an authenticated report from \prv to determine if the correct software is installed on \prv. While effective in detecting malicious code modifications, \RA is oblivious to attacks such as control flow hijacking~\cite{schuster2015counterfeit} that alter the program's behavior without changing instructions. 
Control Flow Integrity (CFI)~\cite{cfi-care,ge2017griffin,Abadi2009} can be used to locally detect some of these attacks on \prv. However, it provides no evidence of the attack behavior to \vrf.

Control Flow Attestation (\CFA)~\cite{cflat,zeitouni2017atrium,dessouky2018litehax,oat,caulfield2023acfa,traces,recfa,scarr,ari,yadav2023whole} provides \vrf the ability to ascertain both the runtime behavior and integrity of \prv. \CFA extends \RA to record a trace of the control flow path followed during the attested program's execution. This trace is created by logging the destinations of all control flow instructions (e.g., \texttt{call}, \texttt{jump}, or \texttt{ret}) executed. The resulting control flow log (\cflog) is authenticated alongside \prv's installed code (per standard \RA) and sent to \vrf. With \cflog, \vrf can determine whether the attested execution had valid runtime behavior. For more details on CFI, \CFA, as well as their differences and similarities, we refer the reader to the systematization in~\cite{sok_cfa_cfi}.

As \cflog contains all branches taken, its storage and eventual transmission are bottlenecks for \CFA.
Early \CFA techniques~\cite{cflat, dessouky2017fat,zeitouni2017atrium} avoided this by compressing \cflog into a single hash digest by computing a hash-chain of all control flow destinations in \cflog. However, as attested programs become more complex, this approach leads to the well-known path explosion problem~\cite{aliasing}, making verification by \vrf infeasible. Similarly, hash-based approaches do not offer insight into malicious control flows taken. As a consequence, more recent \CFA methods~\cite{dessouky2018litehax,oat,caulfield2023acfa,traces,recfa,scarr,ari,yadav2023whole} tend to log paths \textit{verbatim} aside from simple program-agnostic log optimizations (e.g., replacing simple loops with counters).

The above-mentioned program-agnostic optimizations do not capture application-specific characteristics that can offer further \cflog reductions.
Therefore, recent work proposed application-specific \cflog optimizations. SpecCFA~\cite{speccfa} replaces \vrf-defined high-likelihood control flow sub-paths in \cflog with reserved symbols of reduced size. This allows \vrf to speculate on and configure \prv with a set of expected sequences of control flow transfers within the attested application. As a result, SpecCFA achieves significant reductions in the costs of storing and transmitting \cflog. SpecCFA's optimization strategy depends on the predictability of \prv's execution. However, by focusing solely on sub-path frequency, SpecCFA misses other highly predictable application characteristics, such as redundancy in the representation of \cflog data or the locality of instructions within memory.

Based on the observation above, our premise in the present work is that speculating on these other predictable characteristics could further reduce \cflog. Therefore, we propose REpresentation-aware SPECulative \CFA (\acron), a method (accompanied by corresponding architectural design and implementation) to enable secure \CFA speculation based on two new application-specific characteristics:
\begin{itemize}

\item First, \acron allows \vrf to speculate on the locality of instructions in an attested program. MCU applications are typically statically linked to fixed program memory address ranges. Hence, code addresses often share common prefixes. \acron allows \vrf to speculate on the length of this prefix, grouping \cflog entries by shared prefix. Each prefix is added to \cflog with a special symbol to distinguish it from regular addresses. For subsequent entries sharing the same prefix, only the suffix is logged. When a new address has a different prefix, the new prefix is logged, and the process repeats: only suffixes are logged until the next prefix mismatch. This reduces the size of most \cflog entries, removing redundant data without loss of information. 

\item Second, \acron allows \vrf to speculate on \cflog's data representation itself.
For this, \vrf speculates on a Huffman encoding~\cite{huffman2007method} (e.g., based on previously received \cflog-s for the same code) and sends it to \prv along with a \CFA request. Upon receipt, \acron uses the Huffman encoding to compress \cflog at runtime. This allows \CFA to benefit from Huffman-based compression without placing the burden of computing compression algorithms on the resource-limited \prv.

\end{itemize}

\CFA schemes either rely on Trusted Execution Environments (TEEs)~\cite{cflat,oat,ari,traces,recfa,scarr,neto2023isc} or custom hardware support~\cite{caulfield2023acfa,dessouky2018litehax,zeitouni2017atrium,dessouky2019chase}. While \acron's concept applies to both categories, we implement \acron by modifying SpecCFA's TEE-based implementation~\cite{speccfa}.
This choice is motivated by (1) SpecCFA’s open-source availability and (2) our goal of jointly implementing SpecCFA and RESPEC-CFA to maximize the combined benefits of both methods (their combined benefits are later confirmed by our experiments in Section~\ref{sec:eval_combined}).
Therefore, \acron's prototype inherits SpecCFA's characteristic of targeting ``off-the-shelf'' MCUs with TEE support (specifically, ARM Cortex-M with TrustZone). We evaluate \acron's performance using real-world MCU applications and find that it achieves large \cflog reduction with little runtime overhead. When combined with SpecCFA, \acron achieves up to 99.7\% reduction of \cflog size for the evaluated applications. We also make \acron's prototype publicly available at~\cite{repo}.

\section{Background \& Related Work}

\subsection{MCU Architectures}\label{sec:scope}
\label{sec:mcu}

MCUs are compact processors with CPU, memory, and I/O peripherals built into one low-cost chip. They are typically embedded within larger systems and used for sensing/actuation and real-time responses to stimuli. Additionally, they offer low-power execution modes/idle states until asynchronous interrupt-based processing is triggered. These characteristics make them useful for a variety of settings, including those that require lengthy deployments or minimal energy consumption.

The CPU within an MCU is typically single-core and executes software from physical memory (at ``bare-metal''), i.e., without an MMU to enable virtualization and inter-process isolation. On the lower end of the scale (e.g., 8- or 16-bit CPUs from Microchip AVR~\cite{avr-8-bit-mcus} or TI MSP430~\cite{msp430-16-bit-mcus}), they typically run 1-16 MHz clock frequencies with 4-256 KB FLASH or FRAM memory for instructions and 1-64KB SRAM memory for data. As it relates to security resources, many devices are not equipped with extensive modules. In some cases, they might be equipped with general-purpose Memory Protection Units (MPU), but are limited (e.g., support for three configurable regions only in program memory~\cite{msp430-fram-mpu}),
or other security modules (e.g., Intellectual Property Encapsulation in TI MSP430~\cite{msp430_ipe}).

Slightly more advanced MCUs include ARM Cortex-M MCUs (e.g., ARM Cortex-M33 used for prototyping in this work~\cite{arm-cortex-m33}).
The ARM Cortex-M class of MCUs has 32-bit CPUs that typically range from 48-600 MHz clock frequencies, between 16-2048 KB of FLASH memory, and 4-1400 KB SRAM memory~\cite{arm-cortex-m7,arm-cortex-m0}. They are also equipped with a Wake-up Interrupt Controller (WIC) that enables entering idle states and low-power modes.
The ARM Cortex-M class of MCUs also has more security features, such as stronger MPUs (e.g., supporting up to 8-16 configurable regions over all addressable memory) and the TrustZone security extension (discussed further in Section~\ref{sec:trustzone}). Yet, it lacks MMUs/virtual memory.

\subsection{TrustZone for MCUs}\label{sec:trustzone}
\begin{figure}
    \centering
    \includegraphics[width=0.7\linewidth]{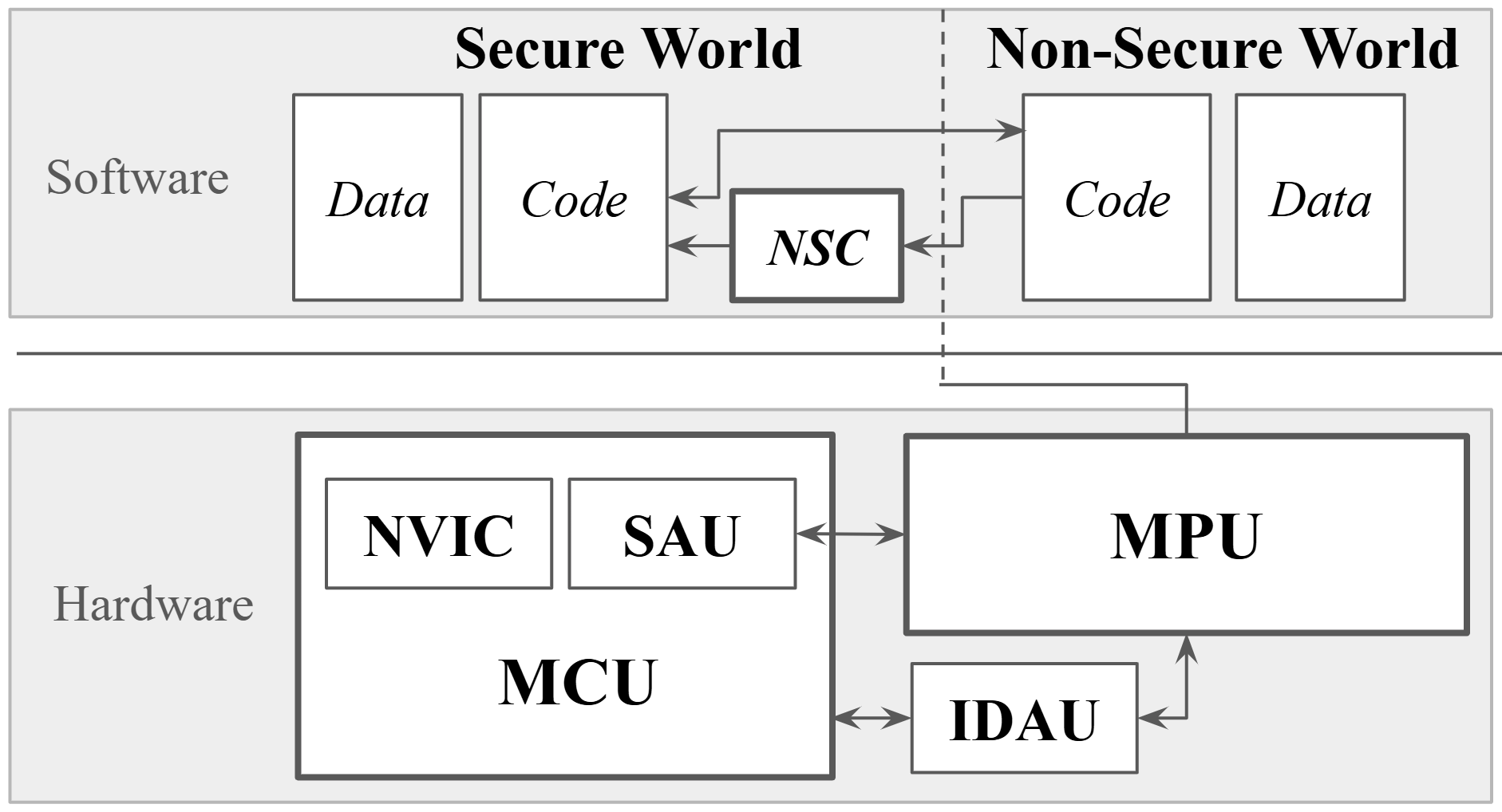}
    \caption{ARM Cortex-M TrustZone}
    \label{fig:trustzone}
\end{figure}

ARMv8 Cortex-M MCUs are equipped with the TrustZone (i.e., TrustZone-M) TEE~\cite{Armv8_M_TZ_spec}, depicted in Figure~\ref{fig:trustzone}. TrustZone provides strong software isolation by dividing hardware and software into two worlds: the ``Non-Secure'' and ``Secure'' worlds.

These worlds are defined by two hardware controllers: the Secure Attribution Unit (SAU) and the Implementation-Defined Attribution Unit (IDAU)~\cite{arm_idau_sau}. 
The region definitions enforced by the IDAU are fixed by the manufacturer, and developers can configure the SAU via the Secure World code to assign additional memory to the Secure World as needed for a particular program. These configurations set by IDAU and SAU are enforced by the MPU alongside any specific inner-world access controls (e.g., setting Non-Secure World code as read and execute only).
Additionally, ARM Cortex-M MCUs are typically equipped with a Nested Vector Interrupt Controller (NVIC)~\cite{arm_nvic_m7,arm_nvic_m33} that manages interrupts. The NVIC can be controlled by Secure World code to assign interrupts to a particular world. It can also be configured to ensure Non-Secure World interrupts do not interrupt the Secure World, and to set Secure World interrupts as higher priority.

TrustZone's hardware-based isolation ensures that the Non-Secure World cannot tamper with code and data belonging to the Secure World~\cite{ARM-Trustzone}. As such, the Secure World can safely store security-critical functionality. 
TrustZone also forces controlled invocation of the Secure World through dedicated entry points called Non-Secure-Callables (NSCs), while enabling the Secure World to call Non-Secure World code directly, as depicted in Figure~\ref{fig:trustzone}.

Prior work has used TrustZone-M to enhance various aspects of embedded system security, including but not limited to availability/performance~\cite{wang2022rt,sbis} and enabling security mechanisms of high-end CPUs (e.g., ALSR without MMUs~\cite{luo2022faslr}, and virtualization~\cite{pinto2019virtualization}). 
Similarly, several works have utilized TrustZone for detecting control flow attacks, whether done locally through CFI~\cite{sherloc,insectacide,cfi-care} or remotely through CFA~\cite{cflat,abera2019diat,oat,ari,neto2023isc,traces}.
For a more comprehensive discussion of TrustZone see~\cite{pinto2019demystifying}.

\subsection{Remote Attestation}\label{sec:ra}
\RA occurs between a \vrf and a potentially compromised \prv, allowing \vrf to remotely assess \prv's state. An \RA instance is comprised of the following core steps (depicted in Figure~\ref{fig:RA}):
\begin{figure}[t]
    \centering
    \scalebox{1}{
    \begin{tikzpicture}
        \fill[blue!40!white] (1,0) rectangle (1.25,2);
        \node[text width=3cm] at (1.75,2.25) {\small Verifier (\vrf)};
        
        \fill[red!40!white] (3.75,0) rectangle (4,2);
        \node[text width=3cm] at (4.75,2.25) {\small Prover (\prv)};

        \draw[->, thick] (1.5, 1.5) -- node[above] {\small (1) Request} (3.5,1.5);

        \node[text width=3cm] at (5.6,1) {\small (2) Authenticated Integrity Check};

        \draw[<-, thick] (1.5, 0.5) -- node[above] {\small (3) Report} (3.5, 0.5);

        \node[text width=2.9cm] at (-0.1,0.2) {\small (4) Verify Report};
    \end{tikzpicture}
    }
    \caption{A typical \RA interaction}
    \label{fig:RA}
\end{figure}
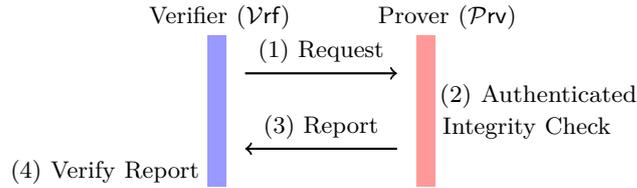
\begin{compactenum}
    \item \vrf sends a cryptographic challenge $Chal$, requesting \prv attest to its current state.
    \item Upon receiving $Chal$, \prv produces a token $H$ by computing an authenticated integrity check on its memory and $Chal$.
    \item \prv responds to \vrf by sending $H$.
    \item \vrf compares $H$ against its expected value to determine if \prv has been compromised.
\end{compactenum}

The authenticated integrity check in step 2 is implemented using a message authentication code (MAC) or a digital signature. Hence, the secret key used to produce $H$ must be securely stored and used by a root of trust (RoT) on \prv in full isolation from any compromised software on \prv. Optionally, the RoT in \prv may also authenticate \vrf requests (in step 1). This mitigates denial-of-service attempts via bogus \RA requests~\cite{ra_prvs_perspective} and ensures that any other data within the request (e.g., \vrf-issued commands in security services that build upon \RA, such as~\cite{caulfield2023acfa,traces,rata,pfb,casu}) are genuine.
In the context of this work, it also ensures that \vrf-defined speculations are authentic.


\RA is generally classified by its RoT implementation. Early schemes relied solely on software to attest the \prv's state. While applicable to commodity MCUs, these software-only approaches require deterministic timing characteristics such as a wired interface between \vrf and \prv for predictable network latency~\cite{swatt,pioneer,sake}. These assumptions often make software-based approaches inapplicable to remote settings~\cite{SW-Attsnt-Attacks}.

Other models~\cite{checkmate_att,tpm_attest,Sancus17} use dedicated hardware and hardware-protected secrets to attest the \prv. Hardware-based approaches provide stronger security guarantees, but the additional hardware cost can be prohibitive for resource-constrained MCUs.

Finally, some \RA schemes~\cite{vrased,smart,tytan} try to find a balance between hardware's strong security guarantees and software's lower cost. These ``hybrid'' approaches typically implement the MAC/signature generation in software while using minimal hardware to securely store the secret key and protect the execution of the RoT software.

\subsection{Control Flow Attestation}\label{sec:cfa}

Control flow attacks alter the execution of a program without modifying code~\cite{szekeres2013sok}. As a result, \RA alone cannot detect these attacks. \CFA extends \RA to generate a \cflog of the attested program's execution by recording the execution of branching instructions (e.g., \texttt{jump}, \texttt{call}, \texttt{ret} instructions) at runtime. To detect these branching instructions and securely store \cflog, existing \CFA techniques rely on either (1) binary instrumentation with TEE-support~\cite{cflat,oat,ari,traces,neto2023isc,scarr,recfa,yadav2023whole} or (2) custom hardware elements~\cite{zeitouni2017atrium,dessouky2017fat,dessouky2018litehax,caulfield2023acfa}. When the attested execution completes, the resulting \cflog is signed/MAC-ed alongside \prv's program memory content (as per \RA) to produce H. Both H and \cflog are sent to \vrf. With H \vrf can validate \prv's code integrity and authenticate \cflog. \cflog tells \vrf the executed path.


Early \CFA schemes used a single hash to represent \cflog~\cite{cflat,zeitouni2017atrium,dessouky2017fat}, compressing the execution into a small fixed-size value. This approach minimized the storage and transmission overhead associated with \CFA. Similarly, to verify a given execution, \vrf simply needs to check if the received hash exists in the set of all valid execution hashes. However, as binaries get more complex, trying to enumerate all possible paths through the program becomes exponentially complex, leading to the path explosion problem~\cite{aliasing}. Further, hash-based approaches can only detect if a given run is invalid. While malicious control flows impact the final hash, the malicious path itself is not visible to \vrf. As a result, \vrf cannot learn what triggered the attack nor how to correct it.


To address these limitations, recent \CFA techniques log all control flow transfers \textit{verbatim}~\cite{dessouky2018litehax,oat,traces,caulfield2023acfa,scarr,recfa,yadav2023whole}. This eases verification; however, \textit{verbatim} logs can quickly outgrow the memory available on MCUs. Hence, prior work introduced several simple \cflog optimizations. Some approaches reduce the size of \cflog by limiting their scope to a subset of operations, such as indirect branches~\cite{nunes2021tiny,oat,nunes2021dialed} or a subset of the code~\cite{ari}. Others reduce the size of log entries themselves rather than the number of entries logged. LiteHAX~\cite{dessouky2018litehax} records conditional branches with a single bit ('1' if the branch was taken, '0' otherwise) while indirect branches are logged in full. OAT~\cite{oat} uses a similar bitstream representation to LiteHAX; however, OAT creates a hash-chain of return addresses rather than logging them directly. Despite using hash-chains, the added context of the rest of \cflog allows OAT to avoid the issues associated with the early hash-based \CFA approaches. Many \CFA techniques also replace repeated loop entries with a count denoting how many times the loop executed~\cite{cflat,caulfield2023acfa,traces,nunes2021tiny,nunes2021dialed,zeitouni2017atrium,recfa}.

Regardless of these optimizations, it is still possible for \cflog to outgrow the available memory. In response, some \CFA controls fix the size of \cflog in memory and transmit the log in slices throughout the attested execution when available memory is full~\cite{scarr,caulfield2023acfa,traces}. On its own, this approach trades storage overhead for increased transmission/runtime costs due to the additional intermediate log transmissions. As such, \CFA techniques often combine this approach with other optimizations to reduce the number of \cflog slices that must be transmitted.

The above-mentioned methods are based on static characteristics common to all programs. As a result, these schemes inherently miss application-specific characteristics that can be leveraged to further reduce \cflog. SpecCFA~\cite{speccfa} demonstrates the benefits of application-aware optimization by allowing \vrf to speculate on high-likelihood control flow sub-paths. From the binary or previously received \cflog-s, \vrf can configure \prv with a set of expected frequently occurring execution paths (e.g., frequent control loops, sensing operations, etc.). At \cflog construction time, SpecCFA replaces these sub-paths in \cflog with small symbols (sub-path IDs), substantially reducing the size of \cflog. As \vrf knows the unique path-to-ID correspondence, this optimization does not result in any loss of information in \cflog.

\section{\acron}

\subsection{Intended Contribution}

In this work, we propose that \vrf can speculate on the data representation characteristics of the attested application to optimize \cflog size during its construction. For instance, MCU applications are typically statically linked within a fixed address range, and branch instructions often use relative offsets, making destination addresses predictable. This results in program locality—common address prefixes—that can be anticipated. Additionally, \cflog data may exhibit skewed distributions due to frequent patterns like loop counters, sub-paths, or commonly accessed address ranges.

Building on these observations, we present a method (and supporting design) that enables \vrf to speculate on address prefix sizes and Huffman encodings tailored to the expected \cflog data. This improves \cflog compression at its construction time. We realize \acron as a TrustZone Secure World module that extends \CFA to support these two key optimizations. We also show how \acron can be composed with the state-of-the-art and the benefits of this composition.

\textit{\textbf{Remark.} Key to \acron's practicality is not burdening resource-constrained \prv with Huffman encoding computation. Instead, \vrf takes on this burden by speculating on the ideal encoding based on previously received \cflog-s. This enables both: reduced \cflog size and minimal runtime overhead on \prv.}

\subsection{System and Adversary Models}\label{sec:sys_adv_models}

\acron targets single-core, bare-metal MCUs (recall Section~\ref{sec:scope}) equipped with TEEs (TrustZone-M, in our prototype). Attested applications (\app-s) execute in the Non-Secure World. The Secure World is used to house trusted software modules, including \acron.
TEE support is used for:
\begin{itemize}
    \item Secure storage of attestation keys, which must be securely provisioned prior to deployment;
    \item Isolation of the Secure World's code and data from any \app in the Non-Secure World;
\end{itemize}
These characteristics can be achieved through standard ARM TrustZone-M v8 architectural support~\cite{Armv8_M_TZ_spec}.

We consider an adversary (\adv) capable of fully compromising \prv's Non-Secure World. \adv can exploit memory vulnerabilities in \prv to perform control flow hijacking or code-reuse attacks. In addition, \adv can manipulate Non-Secure World interrupts and their interrupt service routines (ISRs). \adv cannot modify any Secure World code and data due to the underlying TEE hardware protections. \adv cannot bypass \prv's hardware-enforced controls. TEE-based \CFA relies on binary instrumentation to log control flow transfers. Thus, the code of the application being attested must be immutable during its execution and in between execution and measurement by the underlying \CFA method. This is a standard requirement enforced by various \CFA schemes~\cite{sok_cfa_cfi}.

\subsection{\acron High-Level Workflow}\label{sec:workflow}

\begin{figure*}[t]
    \centering
    \includegraphics[width=0.8\columnwidth]{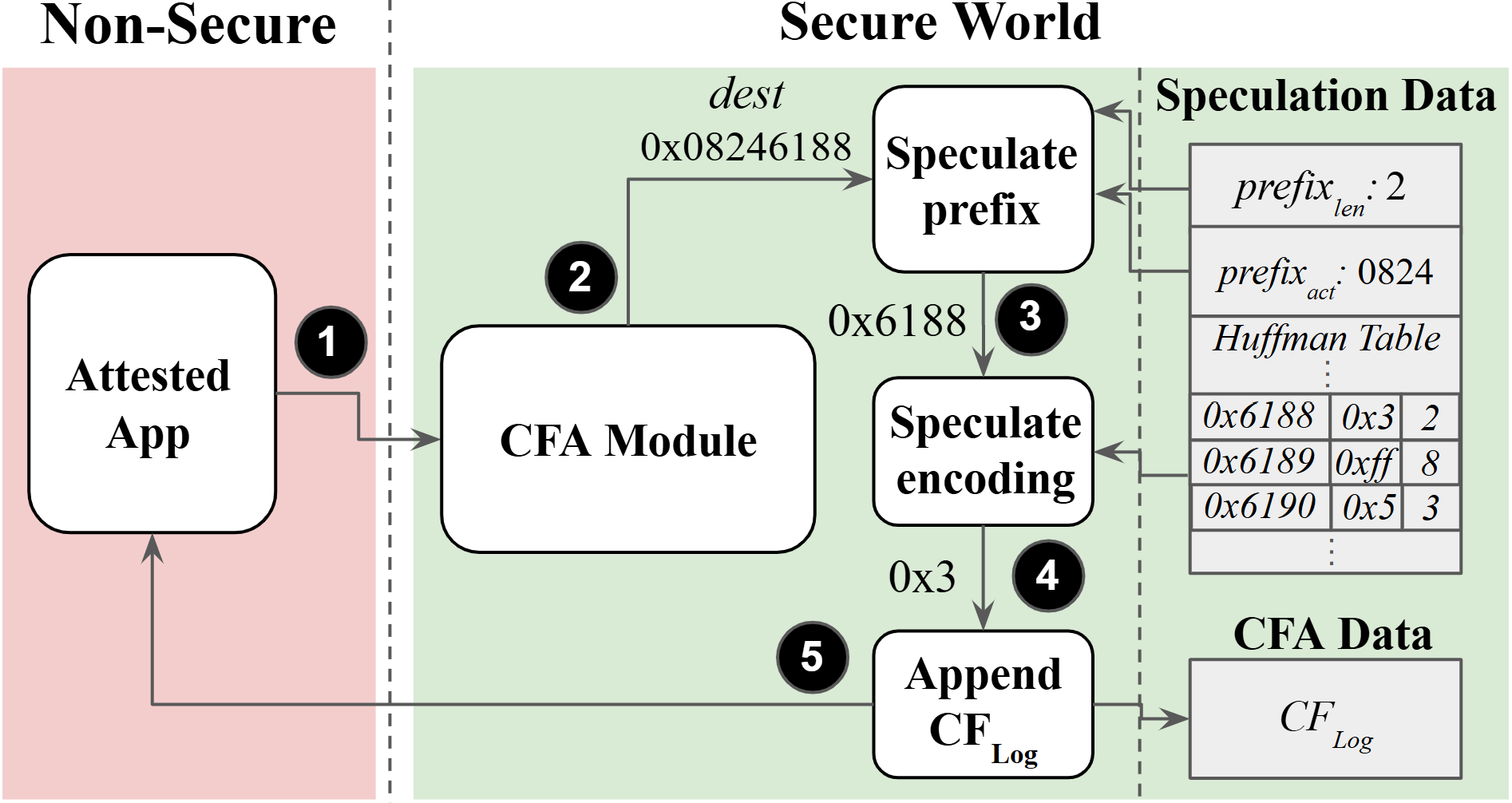}
    \caption{\acron architecture}
    \label{fig:workflow}
\end{figure*}

\acron workflow is shown in Figure~\ref{fig:workflow}.
%
%
To configure \acron, \vrf extends the \CFA request to include a speculated Huffman encoding table and speculated prefix length generated for the attested application \app.
Recall that the request (and the speculation strategy within) are authenticated.
If no speculation is specified in the request, \acron uses previously configured speculations by default.

Upon receiving and authenticating the request, \prv saves the speculated Huffman table and prefix length to Secure World memory and begins \app attested execution in the Non-Secure World. Before deployment, \app is instrumented (as in prior work~\cite{cflat,abera2019diat,neto2023isc,traces,scarr,oat}) with NSC calls to the Secure World at each branching instruction. When each of these instrumented calls executes (step \circled{1}), execution switches to the trusted \CFA module in the Secure World to log the branch destination. The destination address (\textit{dest}) is passed to \acron's first submodule (step \circled{2}). 
In this example, \textit{dest} is the address \texttt{0x08246188}.

\acron's first submodule uses \vrf-configured prefix byte length (\len in Figure~\ref{fig:workflow}). This submodule compares the prefix of \textit{dest} to the current active prefix (\prefix in Figure~\ref{fig:workflow}). As \textit{dest}'s prefix matches \prefix, it is removed from \textit{dest} and the remaining bytes are passed to \acron's next submodule. In this example, the suffix \texttt{0x6188} is given as output in step \circled{3}.

\acron's second submodule uses the \vrf-configured Huffman encoding to compress the suffix; this submodule converts the received data to its corresponding encoding(s). In this example, \texttt{0x6188} maps to the 2-bit Huffman encoding \texttt{0x3}. 
Therefore, in step \circled{4}, the submodule outputs \texttt{0x3} as the final value to be appended to \cflog. After appending \cflog, \acron resumes the execution of \app in step \circled{5}.

The following sections explain the different stages of this workflow in more detail.

\subsection{Prefix Size Speculation Details}\label{sec:prefix}

\acron leverages the locality of MCU software to reduce \cflog's size. Recall from Section~\ref{sec:mcu} that low-end MCUs are typically equipped with limited-sized program memory (e.g., 4 to 2048KB). Within that memory, the attested application generally only makes up a small dedicated portion of it. Further, as attested software is normally statically linked (using custom linker scripts) it has a fixed memory location~\cite{armv8-m-static-linking}. Therefore, \vrf has some prior knowledge of the attested application's memory bounds. Similarly, while some branch instructions can target arbitrary addresses (e.g., indirect jumps), most branch instructions either target a fixed memory address (e.g., direct jumps) or an offset (e.g., conditional branches)~\cite{armv8m-branch-instrs}. Considering these characteristics, it is likely that branch instructions within an attested application visit destination addresses that share some locality. Thus, it is likely that subsequent \cflog entries share a common memory address prefix.

To leverage this, \acron enables \vrf to speculate on the best prefix size to use based on knowledge of the attested application's placement in program memory or analysis of a prior \cflog.
Upon receiving the \CFA request, \acron saves the received prefix length (\len) to Secure World memory. For the first \cflog entry, \acron saves the entry's prefix as the current active prefix (\prefix). \acron then logs the prefix alongside a reserved symbol to indicate to \vrf that this entry denotes a new prefix. After that, \acron adds the entry's suffix to the log. For each subsequent \cflog entry, \acron compares the new entry's prefix to \prefix. If the prefixes match, only the entry's suffix is added to \cflog. Otherwise, the entry's prefix becomes the new \prefix, the new prefix is added to \cflog alongside the reserved prefix symbol, and the entry's suffix is added to \cflog.  

\begin{figure}[t]
    \centering
    \includegraphics[width=\linewidth]{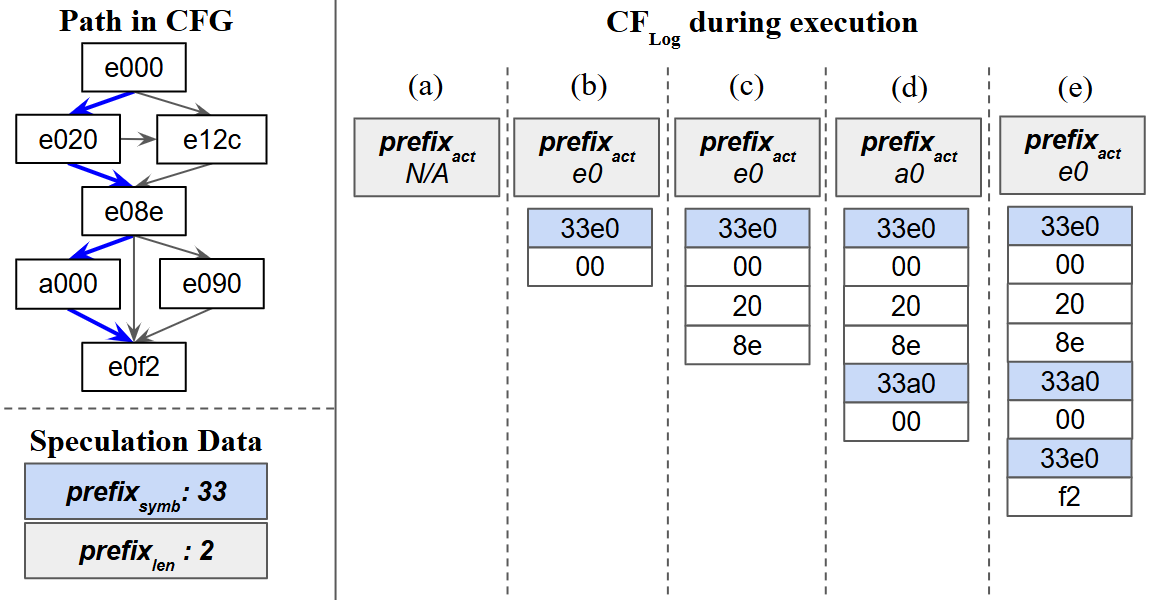}
    \caption{Example \cflog reduction due to prefix size speculation}
    \label{fig:prefix-example}
\end{figure}

A demonstration of the resulting \cflog due to prefix speculation is shown in Figure~\ref{fig:prefix-example}. For the sake of simplicity, this example demonstrates a Control Flow Graph (CFG) with seven nodes, each having a 16-bit start address. In this example, \vrf has selected \len of 2 and configured the reserved prefix symbol as \texttt{33}. When execution starts in (a), no \prefix has been determined yet. The first address is used to select the current \prefix, which is \texttt{e0}. As such, the reserved prefix symbol is logged with \prefix and then the suffix is subsequently logged. Since addresses of the same prefix are visited in (c), only their suffixes are logged. The prefix changes in consecutive control flow transitions in (d) and (e), and thus in both cases, \prefix is updated, the prefix symbol is logged with the new \prefix, and the suffix is logged.

{\bf Note:} If used jointly with other \CFA optimizations that take place before \acron (e.g., loop counters~\cite{cflat} or SpecCFA~\cite{speccfa}), \acron's prefix sub-module might receive non-address inputs (i.e., already optimized entries that do not directly correspond to destination addresses). Non-address inputs are usually encoded with special symbols~\cite{caulfield2023acfa,speccfa}. Therefore, \acron first determines if the input is an address that needs prefix speculation or a special symbol. In the latter, \acron logs a compressed version of the non-address without changing \prefix. 

\subsection{Huffman Encoding Speculation}\label{sec:huffman}

\acron also enables the optimization of \cflog using speculated Huffman encodings~\cite{huffman2007method}. Huffman encoding replaces fixed-length symbols with variable-length codes. The length of these codes is determined by the frequency of symbols in the data, where the more frequently a symbol occurs, the smaller its resulting code. We chose the Huffman algorithm given its optimal encoding properties~\cite{lelewer1987data, huffman2007method}. Nonetheless, we note that any other lossless data encoding scheme of \vrf's choice can also be used. \vrf generates Huffman codes from prior \cflog-s and sends the resulting encoding table to \prv as part of the \CFA request. Further, as new \cflog-s become available, \vrf can use \CFA requests to update the encoding table as desired. \acron uses the received encoding table to convert \cflog entries to their corresponding Huffman code at runtime. The Huffman encoding table is stored in the Secure World on \prv and protected from tampering by \adv. Note that \vrf does not need to send an encoding table with every \CFA request. If no new encoding table is received, \acron continues to use the existing table to encode log entries.

\begin{figure}[t]
    \centering
    \includegraphics[width=\linewidth]{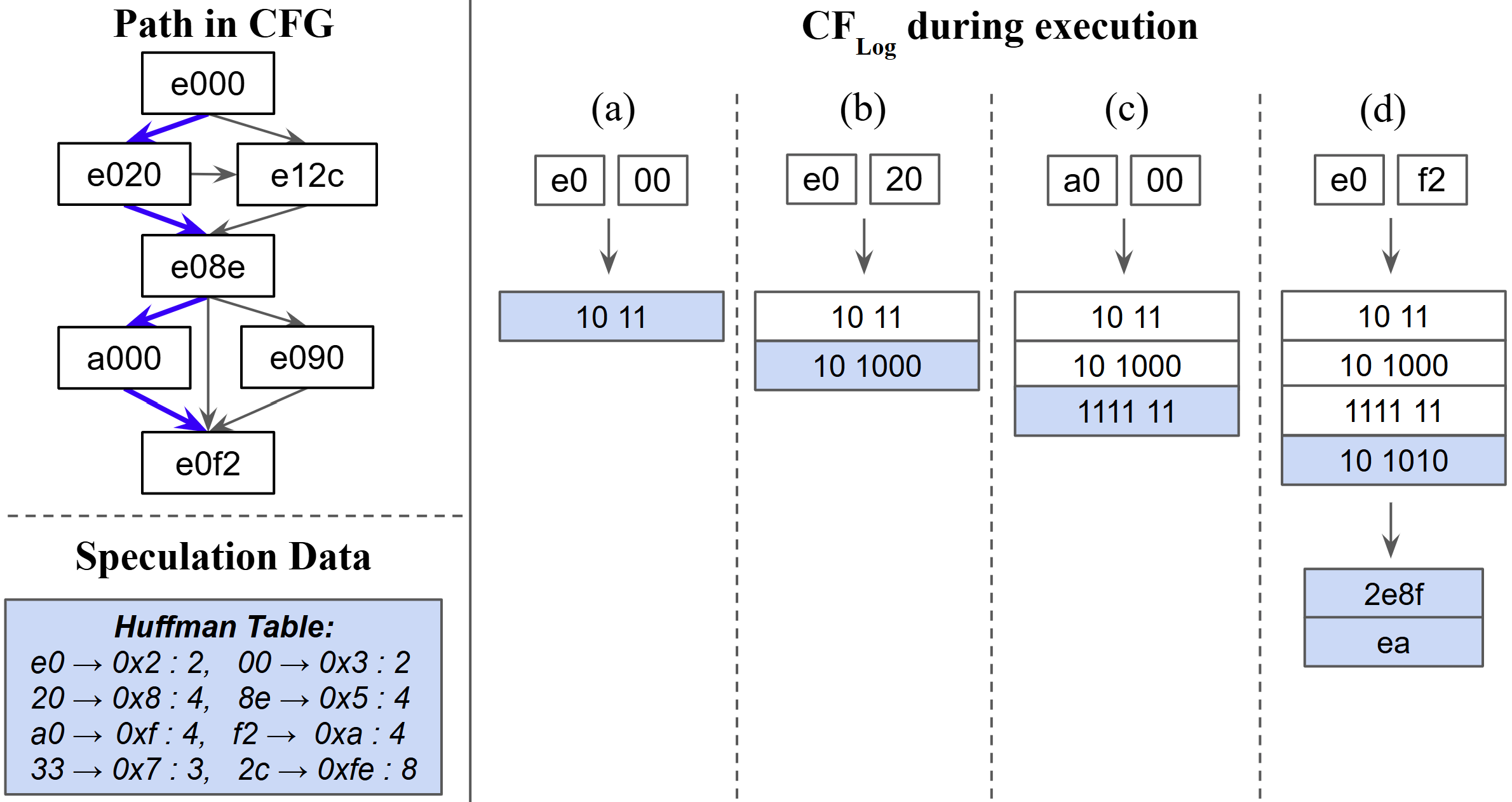}
    \caption{Example \cflog reduction due to Huffman encoding speculation}
    \label{fig:huffman-example}
\end{figure}

An example demonstrating the effect of Huffman encoding speculation is shown in Figure~\ref{fig:huffman-example}. The CFG of \app is the same as the prior example in Figure~\ref{fig:prefix-example}, but now \vrf is configured with a Huffman table denoting the mapping from word to encoding, including the bit length of the encoding. In (a), the first address \texttt{e000} is encoded using the table into its bits into \texttt{1011}. This behavior repeats for each control flow transition in (b)-(d). The final \cflog is represented with the hex values at the end of (d), showing a compressed 3-byte \cflog.



\subsection{\acron Verifier}\label{sec:respec-vrf}
\vrf role includes two additional tasks when \acron is in use. Prior to \prv execution, \vrf generates the speculated Huffman encoding and prefix length. These components are sent to \prv with the \CFA challenge in the initial request. At verification time, \vrf must perform one additional step: decoding of the optimized \cflog into the verbatim \cflog. Naturally, \vrf decodes it by executing the inverse of \prv's encoding steps (shown in Figure~\ref{fig:workflow}). \vrf first uses a locally stored copy of the Huffman encoding table to reverse the encoding in \cflog. Then, it reconstructs the remaining addresses based on the configured prefix length. After that, \vrf performs \cflog verification normally.

\begin{figure*}[h!]
	\abovedisplayskip=0pt
	\abovedisplayshortskip=0pt
	\belowdisplayskip=0pt
	\belowdisplayshortskip=0pt
\centering
\scriptsize
\begin{tabular}{|p{.39\textwidth} p{.15\textwidth} p{.41\textwidth}|} 
\hline
\multicolumn{1}{|c}{\textbf{Verifier} (\vrf)} & & \multicolumn{1}{c|}{\textbf{Prover} (\prv)} \\
\hline

& & \\

1) \vrf generates \CFA challenge (\chal):
\begin{equation*}
\chal \gets \chal_{prev} + 1
\end{equation*}
& & \\

2) \vrf generates speculated Huffman table ($D_{HT}$) using its set of prior \cflog-s ($\mathcal{C}$)):
\begin{equation*}
D_{HT} = \tt{HuffmanEncoding(\mathcal{C})}
\end{equation*}
& & \\

3) \vrf selects speculated prefix length ($prefix_{len}$) & & \\

4) \vrf produces authentication token ($\sigma_{\vrf}$) over the request data: 
\begin{equation*}
\sigma_{\vrf} = \cfattest_\attkey(\chal, D_{HT}, prefix_{len})
\end{equation*} & & \\

5) Create and send \tt{REQUEST}:
\begin{equation*}
\tt{REQUEST} = \{\chal, D_{HT}, prefix_{len}, \sigma_{vrf}\}
\end{equation*}
& 
\sendmessageright{top={\tt{REQUEST}},length={\linewidth}}
&
6) \prv Secure World verifies $\sigma_{\vrf}$ is from \vrf:
\begin{equation*}
	pass := \vrfy_\attkey(\sigma_{\vrf}, \chal, D_{HT}, prefix_{len})
\end{equation*}
And verifies if $\chal$ is valid:
\begin{equation*}
	pass := pass \land (\chal > \chal_{prev})
\end{equation*}
If $pass=TRUE$, continue to step 7. If false, \prv waits for another request before restarting from step 6. 
\\

& & 7) \prv Secure World saves $\{\chal, D_{HT}, prefix_{len}$\} in Secure World data memory\\

& & 8) $PMEM$ section designated for storing \app is made immutable (e.g., via TrustZone hardware controllers as in~\cite{neto2023isc,traces}). \prv executes \app in the Non-Secure World. During execution, the \CFA module in the Secure World will be invoked to build \cflog. With each invocation, \acron will optimize \cflog by referencing $D_{HT}$ and $prefix_{len}$ (see Sec.~\ref{sec:huffman} and~\ref{sec:prefix}). \\

& & 9) After execution completes, \prv Secure World is invoked to compute a \CFA report and update $\chal_{prev}$ ($PMEM$ section designated for storing \app can be made mutable after this stage):
\begin{equation*}
	H := \cfattest_\attkey(\chal, PMEM, \cflog)
\end{equation*}
\begin{equation*}
        \chal_{prev} := \chal
\end{equation*} \\

11) Verify $H$ was produced by \prv:
& 
\sendmessageleft{top={\tt{REPORT}},length={\linewidth}}
& 
10) Construct the \CFA report (\tt{REPORT}): 
\begin{equation*}
	\tt{REPORT} := \{H, \cflog\}
\end{equation*}\\
%
\begin{equation*}
\vrfy(H, \attkey, \chal, PMEM', \cflog)
\end{equation*} & & \\
where $PMEM'$ is the expected content of \prv $PMEM$ (including \app at the expected $PMEM$ region). & & \\

12) Use $D_{HT}$ and $prefix_{len}$ to decode the optimized \cflog and obtain the full \cflog ($\cflog^{V}$):
\begin{equation*}
\cflog^{V}:=(D_{HT}, prefix_{len}, \cflog)
\end{equation*}
& & \\

13) Completes verification by analyzing $\cflog^{V}$.
& & \\

\hline
\end{tabular}
\caption{\CFA protocol with \acron.}
\label{fig:prot}
\end{figure*}

\subsection{End-to-End Protocol}\label{sec:prot}

Figure~\ref{fig:prot} details \acron end-to-end protocol. The protocol assumes the following starting state:
\begin{itemize}
    \item \prv correctly implements \acron (including its underlying CFA architecture) within the Secure World and in isolation from the Non-Secure World.
    \item \vrf and \prv share a symmetric key ($\attkey$). An asymmetric version of the protocol can be obtained in the standard way and is omitted for simplicity.
    \item \vrf has a set of prior \cflog-s ($\mathcal{C}$) that it uses to speculate on the next \cflog.
    \item There is a dedicated region of program memory ($PMEM$) within \prv's Non-Secure World where the binary of the attested \app is expected to be installed. Note that if \app instructions are modified or \app is illegally removed from PMEM, this will be detected by \vrf based on the CFA result.
    \item The \vrf-expected code for \app includes CFA instrumentation used to log control flow destination addresses.
    \item \vrf and \prv persistently store $\chal_{prev}$ as a monotonically increasing counter used for authentication. Initially, $\chal_{prev}$ is zero.
\end{itemize}

Steps 1-4 of the protocol describe \vrf's initial steps to create a \CFA request that is sent in step 5. In step 1, \vrf generates an attestation challenge ($\chal$) by incrementing $\chal_{prev}$. In step 2, \vrf uses $\mathcal{C}$ to generate a Huffman encoding in the form of a table ($D_{HT}$) that maps input words to speculated encodings (as described in Section~\ref{sec:huffman}). In step 3, \vrf selects a speculated prefix length ($prefix_{len}$) based on their knowledge of \app and the locality of branch instructions in $PMEM$. In step 4, \vrf authenticates the data that was generated in the previous three steps ($\chal$, $D_{HT}$, $prefix_{len}$) to produce an authentication token ($\sigma_{\vrf}$). In step 5, \vrf creates and sends the request.

Steps 6-10 describe the tasks by \prv's RoT to extract the request data, construct the \CFA evidence, and respond to \vrf. In step 6, \prv's RoT decodes the request into its individual components and verifies the message. This verification occurs by checking:
\begin{enumerate}
    \item if the request is authentic (i.e., $\sigma_{\vrf}$ was generated over \texttt{REQUEST} using $\attkey$)
    \item and if the request is fresh (i.e., $\chal > \chal_{prev}$).
\end{enumerate}
If both checks succeed, \prv's RoT stores the received metadata into the Secure World data memory in step 7. In step 8, \prv's RoT configures the Non-Secure world (e.g., makes relevant $PMEM$ section immutable) and starts executing \app stored in $PMEM$, during which the \CFA RoT will build \cflog and \acron will speculate on logged data by referencing $D_{HT}$ and $prefix_{len}$. After execution completes (or upon a trigger in runtime auditing~\cite{caulfield2023acfa,traces}), \prv's RoT computes the authenticated measurement over $\chal$, $PMEM$, and \cflog to produce an attestation token $H$ (step 9). Finally, in step 10, \prv's RoT constructs and sends the \CFA report.

Upon receiving the report, \vrf performs steps 11-13 to verify the response. In step 11, \vrf receives the report, extracts $H$ and $\cflog$, and first verifies $H$. In this step, \vrf executes $\vrfy$ to check the following:
\begin{itemize}
    \item \prv's evidence is authentic by determining if $H$ was computed using $\attkey$ over \texttt{REPORT};
    \item \prv's evidence corresponds to the current \CFA request, demonstrated through use of $\chal$ as input to the computation of $H$;
    \item \prv has executed \app, demonstrated through checking $PMEM$ used as input for the computation of $H$ matches the expected program memory ($PMEM'$) containing \app at the expected section; 
    \item the reported \cflog was recorded by \prv's \CFA RoT, demonstrated by its use as input for computing $H$.
\end{itemize}
Steps 12-13 pertain to the \cflog verification. In step 12, \vrf reconstructs the complete verbatim \cflog ($\cflog^{V}$) (as described in Section~\ref{sec:respec-vrf}). 
Finally, in step 13, \vrf performs validation of $\cflog^{V}$ itself to determine if the path followed during execution is valid. 

\subsection{Security Analysis}\label{sec:security}

We analyze \acron's security against \adv with capabilities outlined in Section~\ref{sec:sys_adv_models}.
We argue that \acron's additional optimization strategies do not impact the security guarantees of the underlying \CFA architecture. 

Firstly, \adv may attempt to diverge \app's control flow in a way that will not be recorded in \cflog. However, all branch instructions are instrumented to securely record their destination in the Secure World (this is a consequence of the underlying TEE-based \CFA architecture, rather than a \acron-specific feature). Therefore, \adv must first remove instrumented NSC calls that log branch destinations. However, this is prevented by configuring memory controllers to make \app immutable during an active \CFA session~\cite{traces,neto2023isc}. Note that attempts to illegally modify \app code before or after the \CFA instance (when $PMEM$ section containing \app could be mutable) are also detected by \vrf because the \CFA report contains the state of $PMEM$ during the \CFA instance.
%
\adv may try to overwrite \cflog directly to remove evidence of malicious activity. However, \acron stores \cflog in the Secure World, and thus, it is inaccessible to \adv. Before being sent outside the Secure World and over the network to \vrf, \cflog is MAC-ed (or signed, in the asymmetric setting), making tampering detectable. This implies that \adv would need to forge H to correspond to a fake \cflog. However, this is computationally infeasible as long as the secret key is securely stored (in the Secure World) and a cryptographically secure MAC/signature is used to compute H.

\adv could also attempt to abuse \acron's optimizations to hide malicious activity. For example,
\adv could try to corrupt \prefix to log a control flow hijack as originating from a different region of memory. Doing this, \adv could hide the true source of the attack or disguise malicious behavior as benign transfers. Similarly, \adv could corrupt the Huffman encoding table to encode malicious paths as symbols corresponding to benign entries. However, \acron prevents both \prefix and the Huffman encoding table from being tampered with by storing them in the Secure World.

\adv could also attempt to tamper with \acron's implementation itself, altering the code that performs the optimization to use the incorrect encodings, incorrect prefix, or directly write valid entries despite invalid control flow transfers taking place.
However, both \acron's and the underlying \CFA architecture's code are stored in the Secure World. Thus, they are protected from tampering by \adv residing in the Non-Secure World. Further, only the instrumented NSC calls added to the attested application can modify \cflog. These instructions are protected by TrustZone's hardware and have well-defined behavior when invoked. Therefore, they cannot be abused to log incorrect values, change encodings/prefix values, or overwrite existing \cflog entries.

Finally, \adv could attempt to impersonate \vrf and send \prv a malicious \len or Huffman encoding table to shorten/encode malicious entries to benign values. However, this is prevented by ensuring that \prv's RoT authenticates all \vrf requests, as described in Section~\ref{sec:workflow}. 
Additionally, \adv could attempt to replay messages from \vrf to maintain outdated/incorrect encodings or prefix values. However, \vrf is authenticated based on monotonically increasing \chal (as described in Section~\ref{sec:prot}), making replay attacks infeasible.


\section{Implementation \& Evaluation}\label{sec:impl-eval}

We implement \acron on a NUCLEO-L552ZE-Q development board featuring an STM32L552ZE MCU with ARM TrustZone-M support. This development board is based on ARM-Cortex-M33, operating at 110 MHz. A UART-to-USB interface with a baud rate of 38400 is used for communication with \vrf. We develop \acron's prototype by extending SpecCFA's open-source design with support for the new optimization strategies. For evaluation, we use several open-source MCU applications: an Ultrasonic Ranger~\cite{ultra}, a Temperature Sensor~\cite{temp}, a Syringe Pump~\cite{opensyringe}, a GPS implementation~\cite{gps}, a Geiger Counter~\cite{geiger}, and a Mouse~\cite{mouse}.
By default, we configure \acron to speculate on a 2-byte prefix length (i.e., half a memory address). The speculated Huffman encoding is determined by generating a Huffman encoding from prior \cflog-s of the evaluated applications.

We implement \vrf in Python and run it on an Ubuntu 20.04 machine. \vrf functionality is divided into two scripts. The first script generates a Huffman encoding table from prior \cflog-s for a specified alphabet. Our evaluation is based on a 1-byte encoding Huffman alphabet. The second script decodes received \cflog-s into their full form.

\subsection{\cflog Reductions of \acron in Isolation}

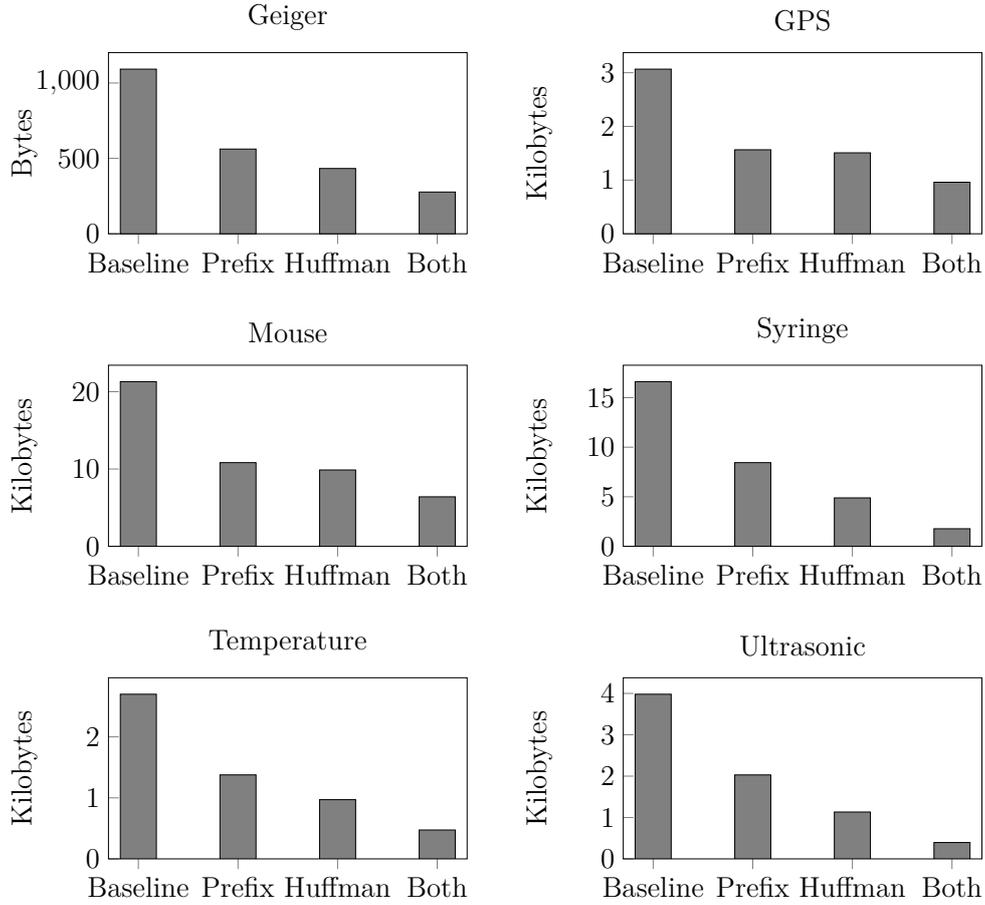
\begin{figure}[t]
    \centering
    \subfigure{
        \scalebox{.9}{
            \begin{tikzpicture}
                \begin{axis}[
                    width=.5\columnwidth,
                    height=4.25cm,
                    ybar,
                    ymin = 0,
                    ylabel = {Bytes},
                    xtick = data,
                    symbolic x coords={Baseline, Prefix, Huffman, Both},
                    title={Geiger},
                    enlarge x limits = .1,
                    label style={font=\large},
                    tick label style={font=\large},
                    title style={font=\large},
                    x label style={at={(axis description cs:0.5,-0.05)}},
                    xtick pos=left,
                    ytick pos=left,
                    bar width = 15pt
                ]
                \addplot[fill=gray] coordinates {
                    (Baseline, 1092)
                    (Prefix, 562)
                    (Huffman, 433)
                    (Both, 277)
                    
                };
                \end{axis}
            \end{tikzpicture}
        }
    }
    \subfigure{
        \scalebox{.9}{
            \begin{tikzpicture}
                \begin{axis}[
                    width=.5\columnwidth,
                    height=4.25cm,
                    ybar,
                    ymin = 0,
                    ylabel = {Kilobytes},
                    xtick = data,
                    symbolic x coords={Baseline, Prefix, Huffman, Both},
                    title={GPS},
                    enlarge x limits = .1,
                    label style={font=\large},
                    tick label style={font=\large},
                    title style={font=\large},
                    x label style={at={(axis description cs:0.5,-0.05)}},
                    xtick pos=left,
                    ytick pos=left,
                    bar width = 15pt
                ]
                \addplot[fill=gray] coordinates {
                    (Baseline, 3.06640625)
                    (Prefix, 1.564453125)
                    (Huffman, 1.5078125)
                    (Both, 0.9599609375)
                    
                };
                \end{axis}
            \end{tikzpicture}
        }
    }
        \subfigure{
        \scalebox{.9}{
            \begin{tikzpicture}
                \begin{axis}[
                    width=.5\columnwidth,
                    height=4.25cm,
                    ybar,
                    ymin = 0,
                    ylabel = {Kilobytes},
                    xtick = data,
                    symbolic x coords={Baseline, Prefix, Huffman, Both},
                    title={Mouse},
                    enlarge x limits = .1,
                    label style={font=\large},
                    tick label style={font=\large},
                    title style={font=\large},
                    x label style={at={(axis description cs:0.5,-0.05)}},
                    xtick pos=left,
                    ytick pos=left,
                    bar width = 15pt
                ]
                \addplot[fill=gray] coordinates {
                    (Baseline, 21.296875)
                    (Prefix, 10.82421875)
                    (Huffman, 9.8779296875)
                    (Both, 6.4052734375)
                    
                };
                \end{axis}
            \end{tikzpicture}
        }
    }
        \subfigure{
        \scalebox{.9}{
            \begin{tikzpicture}
                \begin{axis}[
                    width=.5\columnwidth,
                    height=4.25cm,
                    ybar,
                    ymin = 0,
                    ylabel = {Kilobytes},
                    xtick = data,
                    symbolic x coords={Baseline, Prefix, Huffman, Both},
                    title={Syringe},
                    enlarge x limits = .1,
                    label style={font=\large},
                    tick label style={font=\large},
                    title style={font=\large},
                    x label style={at={(axis description cs:0.5,-0.05)}},
                    xtick pos=left,
                    ytick pos=left,
                    bar width = 15pt
                ]
                \addplot[fill=gray] coordinates {
                    (Baseline, 16.609375)
                    (Prefix, 8.44140625)
                    (Huffman, 4.888671875)
                    (Both, 1.7744140625)
                    
                };
                \end{axis}
            \end{tikzpicture}
        }
    }
        \subfigure{
        \scalebox{.9}{
            \begin{tikzpicture}
                \begin{axis}[
                    width=.5\columnwidth,
                    height=4.25cm,
                    ybar,
                    ymin = 0,
                    ylabel = {Kilobytes},
                    xtick = data,
                    symbolic x coords={Baseline, Prefix, Huffman, Both},
                    title={Temperature},
                    enlarge x limits = .1,
                    label style={font=\large},
                    tick label style={font=\large},
                    title style={font=\large},
                    x label style={at={(axis description cs:0.5,-0.05)}},
                    xtick pos=left,
                    ytick pos=left,
                    bar width = 15pt
                ]
                \addplot[fill=gray] coordinates {
                    (Baseline, 2.69921875)
                    (Prefix, 1.376953125)
                    (Huffman, 0.970703125)
                    (Both, 0.4736328125)
                    
                };
                \end{axis}
            \end{tikzpicture}
        }
    }
        \subfigure{
        \scalebox{.9}{
            \begin{tikzpicture}
                \begin{axis}[
                    width=.5\columnwidth,
                    height=4.25cm,
                    ybar,
                    ymin = 0,
                    ylabel = {Kilobytes},
                    xtick = data,
                    symbolic x coords={Baseline, Prefix, Huffman, Both},
                    title={Ultrasonic},
                    enlarge x limits = .1,
                    label style={font=\large},
                    tick label style={font=\large},
                    title style={font=\large},
                    x label style={at={(axis description cs:0.5,-0.05)}},
                    xtick pos=left,
                    ytick pos=left,
                    bar width = 15pt
                ]
                \addplot[fill=gray] coordinates {
                    (Baseline, 3.98046875)
                    (Prefix, 2.029296875)
                    (Huffman, 1.1328125)
                    (Both, 0.3955078125)
                    
                };
                \end{axis}
            \end{tikzpicture}
        }
    }
    
    \caption{\cflog size: \acron vs. baseline \CFA~\cite{neto2023isc}}
    \label{fig:baseline_comparison}
\end{figure}

We evaluate \acron's impact on \cflog size by comparing \cflog-s generated by a baseline \CFA architecture TRACES~\cite{traces} to \cflog-s generated by the same \CFA architecture equipped with \acron. We evaluate \acron when \vrf has selected to speculate on prefixes alone, Huffman encoding alone, and both. The resulting \cflog sizes for each case are presented in Figure~\ref{fig:baseline_comparison}.

\acron's prefix speculation has a theoretical upper bound based on the size of the prefix compared to the address. Since \acron prototype is built atop ARM Cortex M33 (a 32-bit -- 4 byte -- architecture), configuring \len as 2 bytes results in a theoretical upper bound of \cflog reduction of 50\%. In Figure~\ref{fig:baseline_comparison}, this is observed, as \cflog-s generated by \acron's prefix speculation submodule alone reduce the baseline \cflog-s by 48.5-49.2\%.

\acron's Huffman encoding speculation reduces \cflog by 50.8-71.5\%. Speculating on Huffman encoding is beneficial for programs that change prefixes more frequently, as apparent with the Syringe Pump application in Figure~\ref{fig:baseline_comparison}.

\acron with both strategies achieves the best optimizations, reducing \cflog by 68.7-90.1\%. Since prefixes are optimized away before being processed by the Huffman encoding submodule, the Huffman table can be more fine-tuned to speculate on the encoding of suffixes. Thus, the two submodules complement each other and achieve higher \cflog reductions together.

\subsection{Combined \cflog reductions of \acron + SpecCFA~\cite{speccfa}}\label{sec:eval_combined}

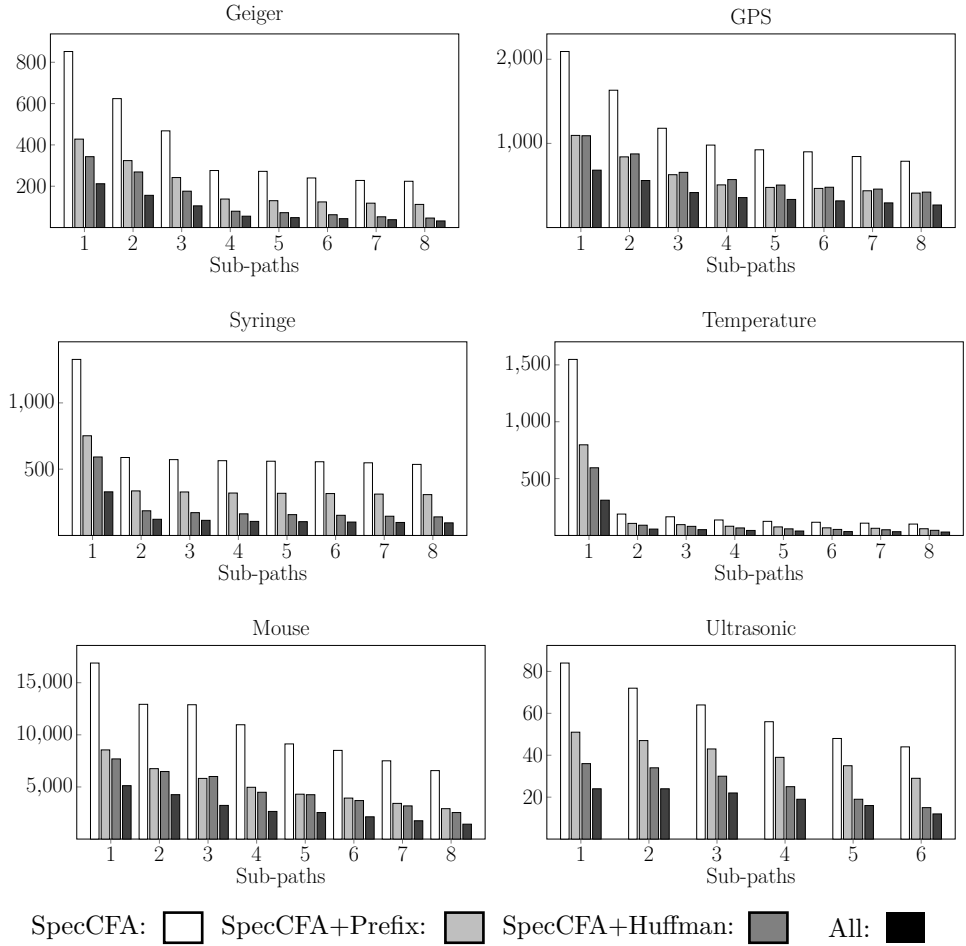
\begin{figure}[t]
    \centering
    \subfigure{
        \scalebox{.4}{
            \begin{tikzpicture}
                \begin{axis}[
                    width=1.1\columnwidth,
                    height=8cm,
                    ybar,
                    ymin = 0,
                    xlabel = {Sub-paths},
                    xtick = data,
                    ytick = {200,400,600,800},
                    title={Geiger},
                    enlarge x limits = .1,
                    label style={font=\huge},
                    tick label style={font=\huge},
                    title style={font=\huge},
                    x label style={at={(axis description cs:0.5,-0.05)}},
                    xtick pos=left,
                    ytick pos=left,
                    bar width = 8pt
                ]
                \addplot[fill=white] coordinates {
                    (1, 852)
                    (2, 624)
                    (3, 468)
                    (4, 276)
                    (5, 272)
                    (6, 240)
                    (7, 228) 
                    (8, 224)
                };
                \addplot[fill=lightgray] coordinates {
                    (1, 428)
                    (2, 324)
                    (3, 242)
                    (4, 138)
                    (5, 130)
                    (6, 124)
                    (7, 118) 
                    (8, 112)
                };
                \addplot[fill=gray] coordinates {
                    (1, 343)
                    (2, 269)
                    (3, 176)
                    (4, 79)
                    (5, 72)
                    (6, 62)
                    (7, 52) 
                    (8, 46)
                };
                \addplot[fill=darkgray] coordinates {
                    (1, 212)
                    (2, 156)
                    (3, 105)
                    (4, 55)
                    (5, 48)
                    (6, 43)
                    (7, 38) 
                    (8, 32)
                };
                \end{axis}
            \end{tikzpicture}
        }
    }
    \subfigure{
        \scalebox{.4}{
            \begin{tikzpicture}
                \begin{axis}[
                    width=1.1\columnwidth,
                    height=8cm,
                    ybar,
                    ymin = 0,
                    xlabel = {Sub-paths},
                    xtick = data,
                    ytick = {1000,2000},
                    title={GPS},
                    enlarge x limits = .1,
                    label style={font=\huge},
                    tick label style={font=\huge},
                    title style={font=\huge},
                    x label style={at={(axis description cs:0.5,-0.05)}},
                    xtick pos=left,
                    ytick pos=left,
                    bar width = 8pt
                ]
                \addplot[fill=white] coordinates {
                    (1, 2092)
                    (2, 1632)
                    (3, 1180)
                    (4, 980)
                    (5, 924)
                    (6, 900)
                    (7, 844) 
                    (8, 788)
                };
                \addplot[fill=lightgray] coordinates {
                    (1, 1095)
                    (2, 840)
                    (3, 628)
                    (4, 507)
                    (5, 477)
                    (6, 465)
                    (7, 437) 
                    (8, 409)
                };
                \addplot[fill=gray] coordinates {
                    (1, 1091)
                    (2, 875)
                    (3, 656)
                    (4, 570)
                    (5, 505)
                    (6, 479)
                    (7, 457) 
                    (8, 421)
                };
                \addplot[fill=darkgray] coordinates {
                    (1, 682)
                    (2, 559)
                    (3, 416)
                    (4, 356)
                    (5, 334)
                    (6, 317)
                    (7, 293) 
                    (8, 268)
                };
                \end{axis}
            \end{tikzpicture}
        }
    }
    \subfigure{
        \scalebox{.4}{
            \begin{tikzpicture}
                \begin{axis}[
                    width=1.1\columnwidth,
                    height=8cm,
                    ybar,
                    ymin = 0,
                    xlabel = {Sub-paths},
                    xtick = data,
                    ytick = {500,1000},
                    title={Syringe},
                    enlarge x limits = .1,
                    label style={font=\huge},
                    tick label style={font=\huge},
                    title style={font=\huge},
                    x label style={at={(axis description cs:0.5,-0.05)}},
                    xtick pos=left,
                    ytick pos=left,
                    bar width = 8pt
                ]
                \addplot[fill=white] coordinates {
                    (1, 1328)
                    (2, 588)
                    (3, 572)
                    (4, 564)
                    (5, 560)
                    (6, 556)
                    (7, 548) 
                    (8, 536)
                };
                \addplot[fill=lightgray] coordinates {
                    (1, 752)
                    (2, 336)
                    (3, 328)
                    (4, 320)
                    (5, 318)
                    (6, 316)
                    (7, 312) 
                    (8, 308)
                };
                \addplot[fill=gray] coordinates {
                    (1, 592)
                    (2, 186)
                    (3, 172)
                    (4, 163)
                    (5, 157)
                    (6, 152)
                    (7, 145) 
                    (8, 140)
                };
                \addplot[fill=darkgray] coordinates {
                    (1, 329)
                    (2, 122)
                    (3, 114)
                    (4, 106)
                    (5, 104)
                    (6, 101)
                    (7, 98) 
                    (8, 95)
                };
                \end{axis}
            \end{tikzpicture}
        }
    }
    \subfigure{
        \scalebox{.4}{
            \begin{tikzpicture}
                \begin{axis}[
                    width=1.1\columnwidth,
                    height=8cm,
                    ybar,
                    ymin = 0,
                    xlabel = {Sub-paths},
                    xtick = data,
                    ytick = {500,1000,1500},
                    title={Temperature},
                    enlarge x limits = .1,
                    label style={font=\huge},
                    tick label style={font=\huge},
                    title style={font=\huge},
                    x label style={at={(axis description cs:0.5,-0.05)}},
                    xtick pos=left,
                    ytick pos=left,
                    bar width = 8pt
                ]
                \addplot[fill=white] coordinates {
                    (1, 1548)
                    (2, 188)
                    (3, 164)
                    (4, 136)
                    (5, 124)
                    (6, 116)
                    (7, 108) 
                    (8, 100)
                };
                \addplot[fill=lightgray] coordinates {
                    (1, 797)
                    (2, 106)
                    (3, 95)
                    (4, 81)
                    (5, 75)
                    (6, 67)
                    (7, 63) 
                    (8, 59)
                };
                \addplot[fill=gray] coordinates {
                    (1, 595)
                    (2, 90)
                    (3, 80)
                    (4, 66)
                    (5, 58)
                    (6, 53)
                    (7, 50) 
                    (8, 45)
                };
                \addplot[fill=darkgray] coordinates {
                    (1, 310)
                    (2, 56)
                    (3, 51)
                    (4, 44)
                    (5, 39)
                    (6, 34)
                    (7, 33) 
                    (8, 30)
                };
                \end{axis}
            \end{tikzpicture}
        }
    }
    \subfigure{
        \scalebox{.4}{
            \begin{tikzpicture}
                \begin{axis}[
                    width=1.1\columnwidth,
                    height=8cm,
                    ybar,
                    ymin = 0,
                    xlabel = {Sub-paths},
                    xtick = data,
                    ytick = {5000,10000,15000},
                    title={Mouse},
                    enlarge x limits = .1,
                    label style={font=\huge},
                    tick label style={font=\huge},
                    title style={font=\huge},
                    x label style={at={(axis description cs:0.5,-0.05)}},
                    scaled y ticks=false,
                    xtick pos=left,
                    ytick pos=left,
                    bar width = 8pt
                ]
                \addplot[fill=white] coordinates {
                    (1, 16888)
                    (2, 12928)
                    (3, 12888)
                    (4, 10968)
                    (5, 9128)
                    (6, 8508)
                    (7, 7508) 
                    (8, 6568)
                };
                \addplot[fill=lightgray] coordinates {
                    (1, 8554)
                    (2, 6756)
                    (3, 5820)
                    (4, 4968)
                    (5, 4306)
                    (6, 3932)
                    (7, 3424) 
                    (8, 2916)
                };
                \addplot[fill=gray] coordinates {
                    (1, 7690)
                    (2, 6480)
                    (3, 6004)
                    (4, 4490)
                    (5, 4256)
                    (6, 3694)
                    (7, 3183) 
                    (8, 2545)
                };
                \addplot[fill=darkgray] coordinates {
                    (1, 5124)
                    (2, 4261)
                    (3, 3230)
                    (4, 2652)
                    (5, 2546)
                    (6, 2137)
                    (7, 1757) 
                    (8, 1430)
                };
                \end{axis}
            \end{tikzpicture}
        }
    }
    \subfigure{
        \scalebox{.4}{
            \begin{tikzpicture}
                \begin{axis}[
                    width=1.1\columnwidth,
                    height=8cm,
                    ybar,
                    ymin = 0,
                    xlabel = {Sub-paths},
                    xtick = data,
                    ytick = {20,40,60,80},
                    title={Ultrasonic},
                    enlarge x limits = .1,
                    label style={font=\huge},
                    tick label style={font=\huge},
                    title style={font=\huge},
                    x label style={at={(axis description cs:0.5,-0.05)}},
                    xtick pos=left,
                    ytick pos=left,
                    bar width = 8pt
                ]
                \addplot[fill=white] coordinates {
                    (1, 84)
                    (2, 72)
                    (3, 64)
                    (4, 56)
                    (5, 48)
                    (6, 44)
                };
                \addplot[fill=lightgray] coordinates {
                    (1, 51)
                    (2, 47)
                    (3, 43)
                    (4, 39)
                    (5, 35)
                    (6, 29)
                };
                \addplot[fill=gray] coordinates {
                    (1, 36)
                    (2, 34)
                    (3, 30)
                    (4, 25)
                    (5, 19)
                    (6, 15)
                };
                \addplot[fill=darkgray] coordinates {
                    (1, 24)
                    (2, 24)
                    (3, 22)
                    (4, 19)
                    (5, 16)
                    (6, 12)
                };
                \end{axis}
            \end{tikzpicture}
        }
    }
    \\
    \begin{tikzpicture}
    \node[text width=1cm] at (0,0) {~};
        
        \node[text width=3cm] at (1.75,0) {SpecCFA:};
        \fill[white] (2,-0.2) rectangle (2.0,0.2);
        \draw[black, line width=1pt] (2,-0.2) rectangle (2.5,0.2);
        
        \node[text width=3cm] at (4.25,0) {SpecCFA+Prefix:};
        \fill[lightgray] (5.7,-0.2) rectangle (6.2,0.2);
        \draw[black, line width=1pt] (5.7,-0.2) rectangle (6.2,0.2);

        \node[text width=3cm] at (7.95,0) {SpecCFA+Huffman:};
        \fill[gray] (9.7,-0.2) rectangle (10.2,0.2);
        \draw[black, line width=1pt] (9.7,-0.2) rectangle (10.2,0.2);

        \node[text width=3cm] at (12.25,0) {All:};
        \fill[black] (11.5,-0.2) rectangle (12.0,0.2);
        
    \end{tikzpicture}
    \caption{Total \cflog bytes after executing each application when \prv is equipped with each speculation strategy.}
    \label{fig:combos}
\end{figure}

To demonstrate \acron's effectiveness alongside existing \CFA speculation strategies, we combine it with SpecCFA and measure the resulting \cflog sizes. To our knowledge, SpecCFA path replacement strategy subsumes the optimizations from prior work and outperforms all other \CFA techniques, making it an ideal candidate for integration and comparison. In this case, \acron workflow (recall Section~\ref{sec:workflow}) takes place after SpecCFA replacement of sub-paths with symbols of reduced size. We evaluate \cflog sizes in the following speculation strategy scenarios: 
\begin{enumerate}
    \item Program sub-path speculation (i.e., SpecCFA) alone;
    \item Program sub-path and \acron's prefix speculation;
    \item Program sub-path and \acron's Huffman encoding speculation; and
    \item All speculation strategies combined (program sub-path speculation from SpecCFA and both prefix and Huffman encoding speculation from \acron) 
\end{enumerate}

By default, SpecCFA supports up to 8 sub-path speculations simultaneously. Therefore, our experiments are also performed varying the number of path speculations from 1 to 8. The results are presented in Figure~\ref{fig:combos}.

Regardless of whether \acron is used in its entirety or partially, it enhances SpecCFA in each of the evaluated cases. \acron's prefix submodule enhances SpecCFA by reducing entries that are not a part of program sub-paths. This is observed in Figure~\ref{fig:combos} by achieving an additional 27.1-55.6\% \cflog reduction from SpecCFA to SpecCFA + prefix. Similarly, \acron's Huffman encoding speculation alone alongside SpecCFA further reduces \cflog sizes by 41.8-79.5\% from SpecCFA-generated \cflog-s.

Finally, the best \cflog reductions are seen when \acron is fully equipped alongside SpecCFA. For the evaluated applications, \acron further reduced SpecCFA \cflog-s by 63.7-85.7\%.
This represents a 91.5-99.7\% reduction in \cflog sizes for different applications, if compared to the baseline \CFA (without any speculation-based strategy), demonstrating synergy in speculating on both \cflog representation and likely sub-paths.

\subsection{Trusted Computing Base (TCB) Size}

\acron's prefix speculation submodule was implemented in 38 lines of C code, and the Huffman encoding speculation submodule was written in 70 lines of code. Additionally, \acron required 26 lines of C code to integrate into SpecCFA. Therefore, \acron in its entirety contributes to a TCB size increase of 134 lines of C code. This correlates to an additional 1140 bytes of Secure World program memory.

\subsection{Memory Overhead}

\acron also requires some Secure World data memory to store the speculation metadata. When speculating on instruction locality, \acron must store the active prefix and its length (1 byte). As a prefix is always shorter than 4 bytes (given ARM Cortex-M 32-bit architecture), the prefix metadata can be stored in at most 5 bytes.

\begin{figure*}
    \centering
    \subfigure[Huffman speculation only]{
        \scalebox{.7}{
        \begin{tikzpicture}
            \begin{axis}[
                width=.65\columnwidth,
                height=5cm,
                ybar,
                ymin = 900,
                ylabel={Bytes},
                xtick = data,
                symbolic x coords={Geiger, GPS, Mouse, Syringe, Temp., Ultra.},
                xtick pos=left,
                ytick pos=left,
                bar width = 15pt
            ]
            \addplot[fill=lightgray] coordinates {
                (Geiger, 972)
                (GPS, 919)
                (Mouse, 947)
                (Syringe, 993)
                (Temp., 987)
                (Ultra., 1000)
            };
            \end{axis}
        \end{tikzpicture}
        \label{huff}
    }
    }
    \subfigure[Huffman \& Prefix speculation enabled]{
        \scalebox{.7}{
        \begin{tikzpicture}
            \begin{axis}[
                width=.65\textwidth,
                height=5cm,
                ybar,
                ymin = 900,
                ylabel = {Bytes},
                xtick = data,
                symbolic x coords={Geiger, GPS, Mouse, Syringe, Temp., Ultra.},
                xtick pos=left,
                ytick pos=left,
                bar width = 15pt
            ]
            \addplot[fill=lightgray] coordinates {
                (Geiger, 970)
                (GPS, 905)
                (Mouse, 928)
                (Syringe, 968)
                (Temp., 987) 
                (Ultra., 998)
            };
            \end{axis}
        \end{tikzpicture}
        \label{huff+prefix}
    }
    }
    \subfigure[Huffman, Prefix, and SpecCFA speculation enabled]{
        \scalebox{.8}{
        \begin{tikzpicture}
            \begin{axis}[
                ybar,
                width=1.2\columnwidth,
                height=5cm,
                ylabel = {Bytes},
                xtick = data,
                symbolic x coords={Geiger, GPS, Mouse, Syringe, Temp., Ultra.},
                xtick pos=left,
                ytick pos=left,
                legend columns=2,
                legend style={at={(.95,0.5)},anchor=west},
                legend image code/.code={
                    \draw [#1] (0cm,-0.1cm) rectangle (0.2cm,0.2cm);
                },
                bar width=5pt
            ]
            \addplot[] coordinates {
                (Geiger, 974)
                (GPS, 904)
                (Mouse, 908)
                (Syringe, 985)
                (Temp., 984)
                (Ultra., 748)
            };
            \addplot[fill=black!20] coordinates {
                (Geiger, 982)
                (GPS, 897)
                (Mouse, 920)
                (Syringe, 988)
                (Temp., 737)
                (Ultra., 747)
            };
            \addplot[fill=gray!80] coordinates {
                (Geiger, 961)
                (GPS, 898)
                (Mouse, 920)
                (Syringe, 991)
                (Temp., 737)
                (Ultra., 749)
            };
            \addplot[fill=darkgray!75] coordinates {
                (Geiger, 747)
                (GPS, 893)
                (Mouse, 932)
                (Syringe, 990)
                (Temp., 738)
                (Ultra., 751)
            };
            \addplot[pattern=crosshatch] coordinates {
                (Geiger, 974)
                (GPS, 894)
                (Mouse, 947)
                (Syringe, 990)
                (Temp., 742)
                (Ultra., 752)
            };
            \addplot[fill=black!20, postaction={pattern=crosshatch}] coordinates {
                (Geiger, 986)
                (GPS, 895)
                (Mouse, 943)
                (Syringe, 992)
                (Temp., 744)
                (Ultra., 757) 
            };
            \addplot[fill=gray!80, postaction={pattern=crosshatch}] coordinates {
                (Geiger, 989)
                (GPS, 901)
                (Mouse, 947)
                (Syringe, 974)
                (Temp., 744)
                (Ultra., 757) 
            };
            \addplot[fill=darkgray!75, postaction={pattern=crosshatch}] coordinates {
                (Geiger, 759)
                (GPS, 908)
                (Mouse, 956)
                (Syringe, 971)
                (Temp., 745)
                (Ultra., 757) 
            };
            \legend{1, 2, 3, 4, 5, 6, 7, 8}
            \end{axis}
        \end{tikzpicture}
        \label{all overhead}
    }
    }
    \caption{Total Huffman table memory overhead on \prv for different applications and \acron configurations}
    \label{huffman overhead}
\end{figure*}
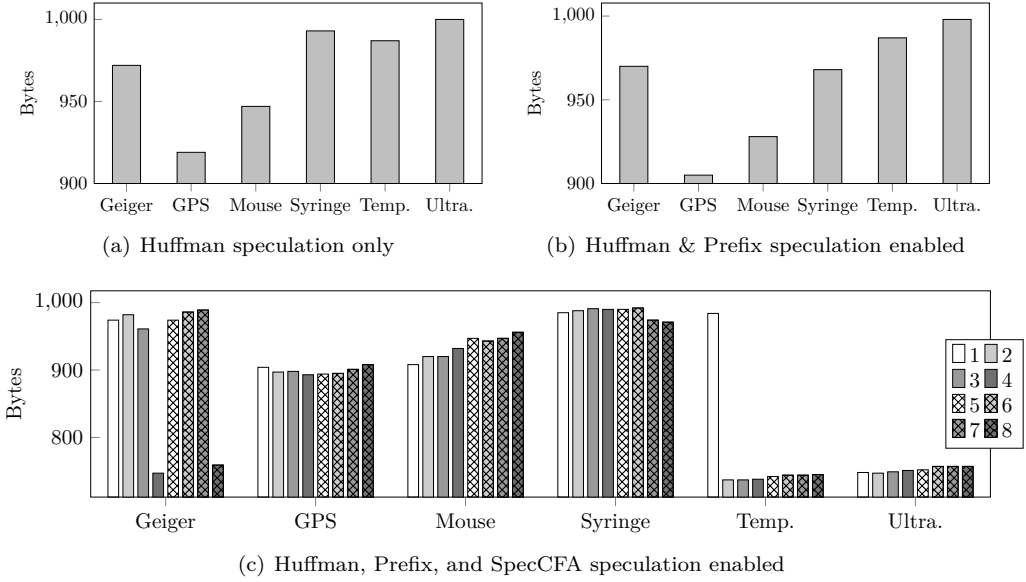

Speculating on Huffman codes has a larger memory impact due to storing the Huffman encoding table.
Figure~\ref{huffman overhead} depicts the total size of the Huffman table for the tested \acron configurations.
In our experiments, we used a 1-byte symbol alphabet to generate Huffman codes, resulting in 256 table entries. 
Each entry is composed of the encoding and its length. The size of Huffman codes varies depending on the attested application and other optimizations enabled (e.g., SpecCFA or \acron's prefix speculation).
Due to this, the total size of Huffman codes ranged from 481 to 744 bytes across all tests.
The length of each code is represented as a single byte, resulting in an additional 256 bytes of overhead. Therefore, when combined, the Huffman table overhead spanned from 737 bytes and 1000 bytes of additional memory overhead in our experiments.

While the size of the Huffman table does vary, the overhead generally is fairly consistent for the evaluated applications, best shown in Figure \ref{all overhead}. However, in some cases, the size of the Huffman table can change drastically. This sudden change in size is due to the relative frequency of data in the \cflog-s used to generate the table.
As mentioned in Section~\ref{sec:huffman}, the more often a symbol appears in the dataset (i.e., a given address in \cflog), the smaller its resulting Huffman code. 
Specifically, Huffman codes are generated using a binary Huffman tree where more frequent symbols are stored higher in the tree~\cite{purdue_huff}. As a consequence, the higher up the tree a symbol appears, the smaller its encoding, but also the less balanced the tree becomes.
Therefore, as the input data becomes more disproportional, so does the length of encodings in the resulting table.
Thus, the Huffman table's size greatly depends on the distribution of entries in \cflog. Changes in input \cflog-s due to other optimizations (e.g., SpecCFA) can greatly alter this distribution, leading to the jumps in Huffman table size seen in Figure~\ref{all overhead}.

\subsection{Runtime Overhead}

\begin{figure*}
    \centering
    \subfigure[Average worst-case time based on total sub-paths configured\label{fig:spec_nsc_time}]{
        \scalebox{.8}{
            \begin{tikzpicture}
                \begin{axis}[
                    width=.6\columnwidth,
                    height=4cm,
                    ybar,
                    ymin = 0,
                    ylabel = {\small Total Time ($\mu$s)},
                    xtick = data,
                    symbolic x coords={B.,1,2,3,4,5,6,7,8},
                    enlarge x limits = .1,
                    label style={font=\Large},
                    tick label style={font=\Large},
                    title style={font=\Large},
                    xtick pos=left,
                    ytick pos=left,
                ]
                \addplot[fill=lightgray] coordinates {
                    (B., 10)
                    (1, 19)
                    (2, 21)
                    (3, 23.2)
                    (4, 25.4)
                    (5, 27)
                    (6, 29) 
                    (7, 32)
                    (8,34.6)
                };
                \end{axis}
            \end{tikzpicture}
        }
    }
    \subfigure[Average worst-case time based on byte length of prefix\label{fig:prefix_nsc_time}]{
        \scalebox{.8}{
            \begin{tikzpicture}
                \begin{axis}[
                    width=.55\columnwidth,
                    height=4cm,
                    ybar,
                    ymin = 0,
                    ylabel = {\small Total Time ($\mu$s)},
                    xtick = data,
                    symbolic x coords={B.,1,2,3,4},
                    enlarge x limits = .1,
                    label style={font=\Large},
                    tick label style={font=\Large},
                    title style={font=\Large},
                    xtick pos=left,
                    ytick pos=left,
                    bar width = 12pt
                ]
                \addplot[fill=lightgray] coordinates {
                    (B., 10)
                    (1, 15.2)
                    (2, 14)
                    (3, 12.6)
                    (4, 11)
                };
                \end{axis}
            \end{tikzpicture}
        }
    }
    \subfigure[Average worst-case time based on bit length of Huffman Encoding\label{fig:huffman_nsc_time}]{
        \scalebox{.6}{
            \begin{tikzpicture}
                \begin{axis}[
                    width=1.5\columnwidth,
                    height=5cm,
                    ybar,
                    ymin = 0,
                    ylabel = {Total Time ($\mu$s)},
                    xtick = data,
                    symbolic x coords={B.,1,2,3,4,5,6,7,8,9,10,11,12,13,14,15,16},
                    enlarge x limits = .05,
                    label style={font=\Large},
                    tick label style={font=\LARGE},
                    title style={font=\LARGE},
                    xtick pos=left,
                    ytick pos=left,
                    bar width = 15pt
                ]
                \addplot[fill=lightgray] coordinates {
                    (B., 10)
                    (1, 29.8)
                    (2, 33)
                    (3, 39.6)
                    (4, 39.8)
                    (5, 49)
                    (6, 52.4)
                    (7, 59.2) 
                    (8, 52.2)
                    (9, 69)
                    (10, 72.2)
                    (11, 78.4)
                    (12, 78.6)
                    (13, 88.8)
                    (14, 92)
                    (15, 98.8) 
                    (16, 92)
                };
                \end{axis}
            \end{tikzpicture}
        }
    }
    \caption{Average worst-case added NSC time per log entry for varying speculation strategies. \texttt{(B. = Baseline)}}
    \label{fig:nsc_time}
\end{figure*}
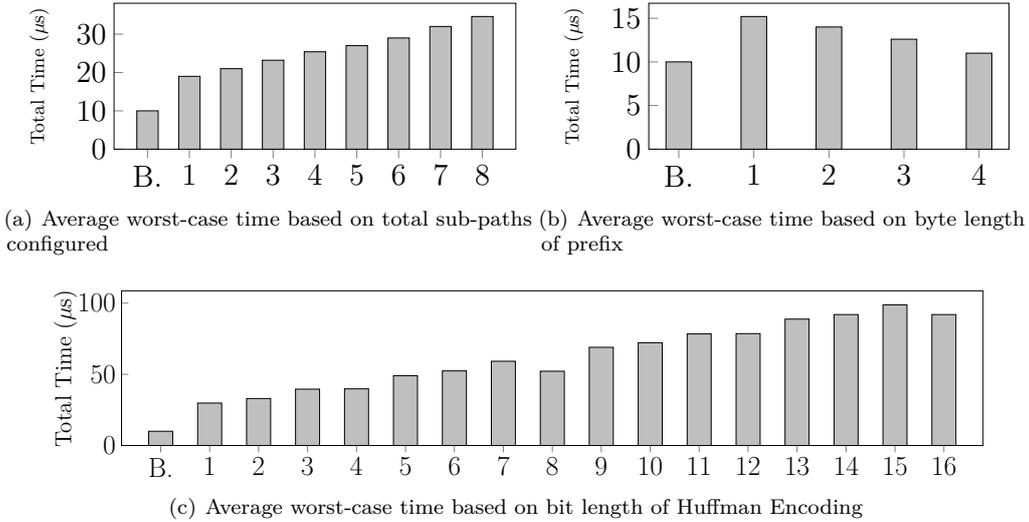

The best \cflog reductions are achieved with \acron and SpecCFA combined. However, the additional submodules added to the Secure World execute upon each NSC. As a result, the time to handle NSCs increases. To evaluate this, in Figure~\ref{fig:nsc_time} we measure the average NSC time to process one entry on applications crafted to target the worst-case timing for each Secure World submodule: SpecCFA, prefix speculation, and Huffman encoding speculation.

Figure~\ref{fig:spec_nsc_time} shows the worst-case time to speculate on sub-paths by SpecCFA, the baseline when \acron extends it. To create the worst-case scenario, \acron varies the total number of sub-path speculations and configures them so all sub-paths mismatch except for the last configured sub-path (i.e., when configured with 8 sub-path speculations, all mismatch except for sub-path 8). In this case, there is an initial $\approx9\mu$s increase from baseline to 1 sub-path. After that, there is a linear increase of $\approx2.22\mu$s per additional sub-path.

Figure~\ref{fig:prefix_nsc_time} shows the worst-case time to speculate on memory address prefixes. For the worst-case application, we craft a program that constantly crosses the configured prefix range. As described in Section~\ref{sec:prefix}, a special ID is logged to denote a change of prefix. Since this ID is the same length as the remaining suffix, \acron suffers more runtime overhead when the prefix is shorter. This is because the ID is longer and is logged more often in this worst-case scenario. However, this scenario is unlikely since \vrf would configure \len based on the anticipated behavior.

Finally, Figure~\ref{fig:huffman_nsc_time} shows the worst-case time to speculate on a Huffman encoding, which occurs when each byte in the address uses the longest bit-length code from the Huffman table. To examine the impact of encoding length, we measure the time for encoding with encoding lengths from 1 to 16 bits. As shown in Figure~\ref{fig:huffman_nsc_time}, the total added time generally increases with the bit length. However, at bit lengths that are multiples of 4, the time improves due to architectural characteristics that enhance the performance on even bytes/half-bytes rather than on uneven bit lengths that do not align in this way.

\section{Discussion}\label{sec:discussion}

\subsection{Worst Case Scenarios.} 
%
The speculation strategies presented rely on prior \cflog-s to generate the appropriate encodings. Thus, a worst-case occurs when no prior \cflog exists yet/is available. Without prior context, neither strategy can accurately predict the application's behavior resulting in no/minimal savings. After obtaining a first \cflog, subsequent speculations can be generated normally. 

The prefix speculation strategy uses a \vrf-defined prefix length to optimize \cflog based on the common locality of branch destinations. If a suboptimal length is chosen (e.g., too long), it is more likely that subsequent \cflog entries will not share a common prefix resulting in more \cflog prefix entries and lower savings. While in theory possible, this scenario is in practice very unlikely due to the simplicity of finding common prefixes in \cflog.

Savings due to Huffman encoding depend on high-frequency symbols in the alphabet. Thus, if symbols are uniformly distributed in \cflog, no savings would occur. 
Similarly, if a particular \cflog has a large number of uncommon symbols, savings gained from the Huffman encoding may be counteracted by the larger encoding of rarer symbols. Fortunately, both scenarios are unlikely due to the type of data in \cflog, i.e., branch destinations that have small cardinality (a subset of program memory's addresses) and occur repetitively.

\acron with Huffman and prefix strategies in combination (or alongside other speculation strategies, such as  SpecCFA \cite{speccfa}) can further reduce the likelihood of the above worst-case scenarios as they cover each other's worst cases. In the case of poor prefixing, each additional prefix entry adds repeated symbols to \cflog. Thus, Huffman encoding would replace these entries with smaller symbols minimizing their impact. Similarly, since prefixing removes repeated portions of memory addresses in \cflog, Huffman encoding can better optimize the remaining symbols. 

Lastly, in some cases, a Huffman table may become larger than the savings it yields in a single \CFA instance. However, since the same table can be reused across multiple \CFA responses, the protocol bandwidth savings grow linearly with the number of protocol instances while the storage cost remains constant. Thus, Huffman encoding is still likely to be cost-effective over multiple instances (i.e., over time).

\subsection{\acron with Interrupts} 
Embedded applications often rely on interrupts for real-time event handling. When an interrupt occurs, the application is paused and execution jumps to an associated ISR to handle the event. Once the ISR is finished, execution returns to the program and the application resumes. Therefore, interrupts affect an application's control flow paths. 

Being agnostic to the underlying \CFA architecture, \acron inherits support for interrupts from the underlying \CFA architecture it builds upon. Some \CFA schemes allow interrupts but do not log them to \cflog~\cite{neto2023isc}. In this case, interrupts do not affect \acron's speculation strategies as they do not appear in \cflog. For architectures that record interrupts~\cite{caulfield2023acfa,scarr,oat}, \acron would speculate on interrupts similar to regular branch addresses in \cflog.

\subsection{\acron in High-End Systems}\label{sec:disc-high-end}
As discussed in Section~\ref{sec:sys_adv_models}, \acron is envisioned for MCUs with limited memory and resources to transmit large \cflog-s. 
Albeit not designed for high-end devices, \acron concepts should also apply in that setting. Larger systems have larger applications and thus more varied \cflog entries. Yet, certain instructions/addresses will still occur more often than others. Therefore, Huffman encoding would still replace high-frequency entries with a shorter code and reduce the size of \cflog. Similarly, prefix speculations would still apply given the locality in execution of software, which occurs in both high-end and low-end devices. Applications in high-end systems are dynamically linked over larger regions of memory, but instructions for different sections of a program (i.e., within a function or library) are typically stored together, making this method applicable.

Regardless of conceptual applicability, in a high-end system, the cost to compute Huffman encodings or determine common prefixes on the fly (or in parallel) might be relatively small or negligible. This would, in turn, obviate the demand for \vrf-based path speculation observed in low-end MCUs.  

\section{Future Directions}
 
\textbf{Static Analysis for Speculation:}
In our current \acron prototype, \vrf generates speculations using \cflog-s from prior executions. While this leads to more optimal speculations, it lacks a mechanism for generating initial speculations when no prior \cflog is available. Future work could address this by developing a static analysis framework that enables \vrf to infer initial speculations from source code and binaries alone. A key challenge is tuning these speculations without knowledge of the actual execution path. This would require techniques that can reason about data representation without prior execution context.

\textbf{\acron in Hardware:}
\acron's design assumes general-purpose TEE hardware support is available on the MCU. However, many \CFA approaches propose custom hardware extensions (as described in Section~\ref{sec:cfa}) to reduce runtime/memory overheads incurred by executing/installing an instrumented Non-Secure world application. Closely related work SpecCFA~\cite{speccfa} proposed a hardware extension and TEE-based variant for their application-specific sub-path speculations. Therefore, future work could include developing a hardware extension to enable \acron in hardware. A challenge will be to determine a representation of the Huffman encoding table that minimizes hardware overhead to make the solution suitable for lower-end devices.

\textbf{Alternative Encodings and Alphabets:}
One component of \acron is the use of Huffman encoding with complete alphabets defined by 1-byte length (e.g., in Section~\ref{sec:impl-eval}, \vrf uses all 1-byte values). A key challenge with larger alphabets is the impracticality of storing the Huffman table on \prv. Future work could explore alternative entropy coding methods -- such as arithmetic encoding~\cite{witten1987arithmetic} -- to support larger alphabets with lower storage overhead. Another direction is to reduce the Huffman input space using domain-specific knowledge about \app, such as valid address ranges/control flow destinations of \app's CFG. However, this raises the issue of how \prv should handle addresses outside the reduced space. The latter may occur during control flow attacks that should also be logged to \cflog.

\textbf{Speculating on Data Flow:}
Another class of runtime attestation is Data Flow Attestation (DFA)~\cite{dessouky2018litehax,nunes2021dialed,abera2019diat} which extends \CFA to also include data flow events in hopes of detecting data-oriented attacks. Data-representation-based speculation proposed in this work might also be suitable for speculating on data flows in MCUs because the bounds of data regions (e.g., ranges for the stack, global data, peripheral memory) are typically fixed and can be determined at compile time.


\section{Conclusion}

We propose \acron to enable speculation and \cflog optimization based on two new properties. First, \acron allows \vrf to speculate on the locality of branch destinations, reducing \cflog size based on shared prefixes across sequences of destinations. Second, \acron enables speculation on the Huffman encoding of \cflog, replacing entries with their corresponding Huffman code at \cflog construction time. We implement an open-source \acron design and evaluate it~\cite{repo}. Our experiments show that \acron results in significant \cflog reductions with little runtime cost. When coupled with prior work in SpecCFA~\cite{speccfa}, further savings are obtained.

\bibliographystyle{alpha}
\bibliography{huffman}

\end{document}